\documentclass[pre,twocolumn,showpacs,showkeys,groupedaddress,longbibliography,preprintnumbers,superscriptaddress]{revtex4-2} 

\usepackage{verbatim}
\usepackage{graphicx}
\usepackage{amsmath}
\usepackage{wrapfig}
\usepackage{epsfig}
\usepackage{float}
\usepackage{amsmath,amssymb,amsfonts,bm}
\usepackage{array}
\usepackage{color}
\usepackage{bbold,ulem}
\usepackage{todonotes}
\usepackage{sidecap}
\usepackage{placeins}
\usepackage{mathtools}
\usepackage{hyperref}
\hypersetup{
	colorlinks=false,
	citebordercolor = lightgray,
	linkcolor=blue,
	filecolor=teal,
	urlcolor=teal,
}
\newcommand{\floor}[1]{\left\lfloor #1 \right\rfloor}
\newcommand{\ceil}[1]{\left\lceil #1 \right\rceil}

\newcommand{\whyCite}[1]{}

\newcommand{\nl}{\sqrt{3}\sigma}
\newcommand{\Lsij}{s^L_{ij}(t)}
\newcommand{\Gsij}{s^G_{ij}(t)}
\newcommand{\Ls}{s^L}
\newcommand{\Gs}{s^G}
\newcommand{\cij}{c_{ij}}
\newcommand{\LNij}{\xi_{ij}^L(t)}
\newcommand{\GNij}{\xi_{ij}^G(t)}
\newcommand{\GN}{\xi^G}
\newcommand{\LN}{\xi^L}
\newcommand{\Gstd}{\sigma^G}
\newcommand{\Lstd}{\sigma^L}

\newcommand{\Grad}{pure gradient}
\renewcommand{\d}{\text{d}}

\usepackage{xr}

\begin{document}

\title{Robust boundary formation in a morphogen gradient via cell-cell signaling}

\author{Mareike Bojer}
\affiliation{Department of Physics, Technische Universit\"at M\"unchen, D-85748 Garching, Germany}

\author{Stephan Kremser}
\affiliation{Department of Physics, Technische Universit\"at M\"unchen, D-85748 Garching, Germany}

\author{Ulrich Gerland}
\altaffiliation{gerland@tum.de}
\affiliation{Department of Physics, Technische Universit\"at M\"unchen, D-85748 Garching, Germany}

\begin{abstract}
Establishing sharp and correctly positioned boundaries in spatial gene expression patterns is a central task in both developmental and synthetic biology. We consider situations where a global morphogen gradient provides positional information to cells but is insufficient to ensure the required boundary precision, due to different types of noise in the system. In a conceptual model, we quantitatively compare three mechanisms, which combine the global signal with local signaling between neighboring cells, to enhance the boundary formation process. These mechanisms differ with respect to the way in which they combine the signals by following either an AND, an OR, or a SUM rule. Within our model, we analyze the dynamics of the boundary formation process, and the fuzziness of the resulting boundary. Furthermore, we consider the tunability of the boundary position, and its scaling with system size.
We find that all three mechanisms produce less fuzzy boundaries than the purely gradient-based reference mechanism, even in the regime of high noise in the local signals relative to the noise in the global signal. Among the three mechanisms, the SUM rule produces the most accurate boundary. However, in contrast to the other two mechanisms, it requires noise to exit metastable states and rapidly reach the stable boundary pattern.
\end{abstract}

\maketitle

\section{Introduction}

The formation and maintenance of gene expression boundaries between neighboring groups of cells is a fundamental task in biology \cite{Dahmann2011, Barkai2020}.
In developmental biology, such boundaries form to spatially partition embryonic tissues into distinct cell fates.
The position of a boundary can be controlled by a global morphogen,
a substance that displays a concentration gradient over the tissue and permits cells to determine their position via a concentration measurement \cite{Wolpert1969}.
Classic questions about gene expression boundaries concern the accuracy and scaling of their positioning \cite{Gregor2007}
and their sharpness \cite{Dahmann2011}.
Embryo-to-embryo variations of the boundary position have been quantitatively studied, for instance, in the vertebrate neural tube \cite{Zagorski2017} and \textit{Drosophila} embryos \cite{Petkova2019}. Studying boundary sharpness additionally requires measuring the boundary profile within each embryo, as was done in Ref.~\cite{Aliee2012} for a mechanically maintained compartment boundary.

Establishing and maintaining sharp,
precisely positioned boundaries that scale with tissue size is a challenging task,
given the various sources of noise in these systems \cite{Lander2013}.
Despite much research, the interplay of mechanisms that performs this task in different organisms and tissues is only partially understood \cite{Bollenbach2008, Jaeger2008, Lander2013, Exelby2021}.
Local cell-cell signaling is well known to play a major role in development \cite{Perrimon2012, Levin2021}, but its role in boundary formation remains underexplored. %
Recently, a quantitative exploration of tissue patterning principles has become feasible in synthetic biology \cite{Barkai2020}.
In a bottom-up approach, synthetic morphogen gradients and different intracellular regulation networks were constructed to quantitatively characterize their interplay \cite{Toda2020, Stapornwongkul2020}.
These studies were able to recapitulate, in a controlled way, many patterning features of native systems,
including boundary formation \cite{Toda2020}.
However, the obtained boundaries were fuzzy rather than sharp, suggesting that additional mechanisms are required to recapitulate this feature of native systems \cite{Barkai2020}.

Here we focus on local cell-cell signaling as a candidate additional mechanism to enable sharp boundaries
and take a theoretical approach to explore its interplay with a global morphogen signal.
A theoretical exploration appears timely, given that engineered cell-cell communication networks have recently been demonstrated \cite{Toda2019},
and further experimental work can benefit from a theoretical comparison of different possible designs.
Also, the same conceptual question arises in a completely different biological context,
based on bacterial systems \cite{Barbier2020}, where short-range signaling can now also be controlled \cite{VanGestel2021}.

Our aim is to identify generic principles rather than to model a specific biological system.
We therefore choose a simple class of models,
permitting us to focus on the conceptual question and to perform a systematic exploration of the associated parameter space.
A prior study in a similar spirit \cite{Hillenbrand2016} analyzed the encoding of positional information in a one-dimensional equilibrium model
(a variant of an Ising spin system), which combined a morphogen gradient with a local interaction between neighboring cells.
Here we focus on the concrete task of boundary formation in a two-dimensional system,
rather than the more abstract notion of positional information, and study both the dynamics and the steady-state properties of the boundary formation process.
Furthermore, we compare three different rules with which cells combine the local and global signals they receive:
the SUM, AND, and OR rule, which are common regulation schemes in biological signal processing \cite{Buchler2003, Mayo2006}.
We investigate which signal processing rule best meets the following criteria in each noise regime:
(i) reduction of the boundary fuzziness,
(ii) short time to reach the stationary boundary position,
(iii) tunability of the boundary to different positions, such that the same mechanism can produce variations of the pattern in related species, and
(iv) scaling of the boundary position with system size, such that pattern proportions are conserved in systems of different sizes.

We find that combining global and local signaling outperforms a gradient-only mechanism in nearly all noise regimes,
even though local signaling adds an additional source of noise to the system.
Among the three different rules, the SUM rule performs best for larger noise levels
but converges slowly to the correct boundary position at lower noise levels.
The performance of the AND and OR rules is equivalent.
The stationary boundary position is tunable within the system for all rules, by varying the threshold for global signaling.
Also, the position scales linearly with system size.

\FloatBarrier
\begin{figure*}[t]
\centering
\includegraphics[trim=0 20cm 0 0, clip,width=0.9\textwidth]{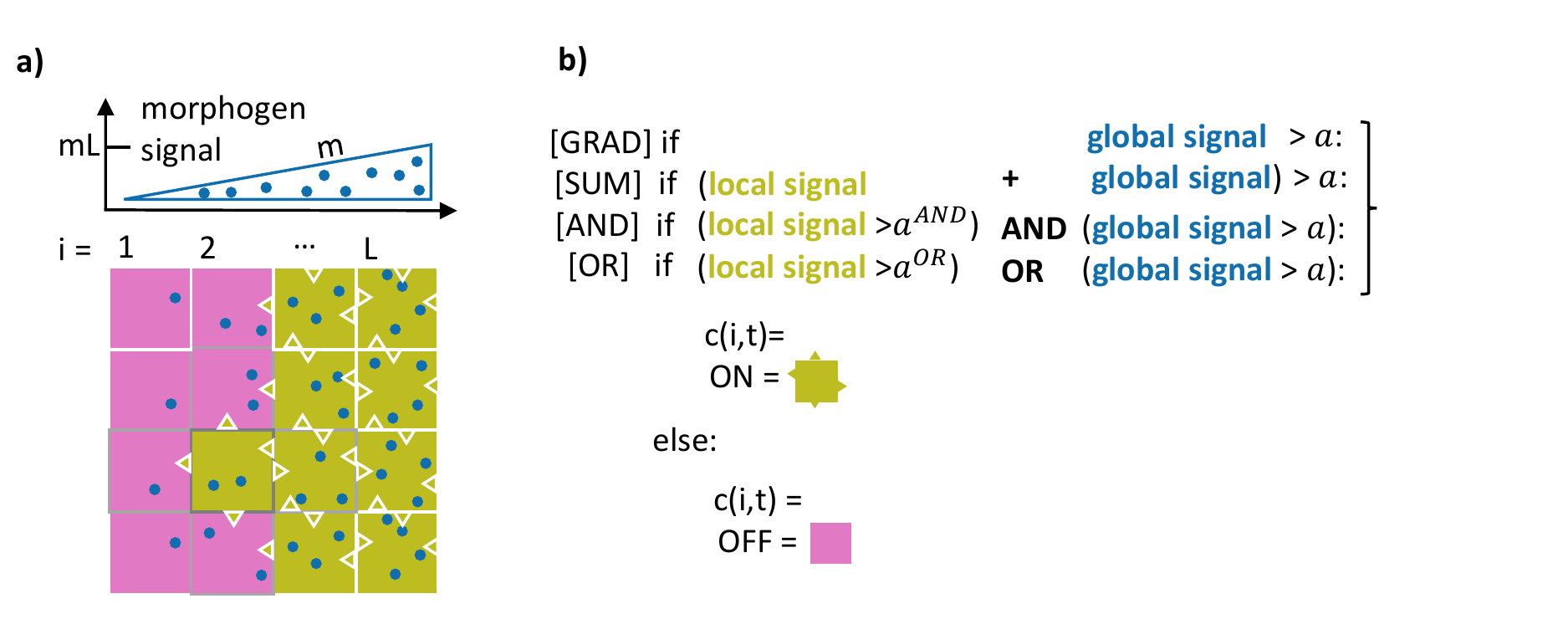}
\caption{
(a) Sketch of the minimal model.
The morphogen signal is represented by blue dots and nearest-neighbor interaction by green triangles.
A green cell is in state On and signals this to its direct neighbors, a pink cell is Off, i.e., not signaling.
The left boundary at $i=0$ is fixed to Off cells, while the right boundary at $i=L+1$ is fixed to On cells.
In perpendicular direction we chose periodic boundary conditions.
For example, the green cell highlighted by a dark gray frame senses the local signal from its upper, lower, left and right neighbor
(light gray frames) canceling to one Off state plus local noise as well as a morphogen concentration of $2m$ plus global noise.
(b) Summary of the three rules:
[SUM]: Given a cell at position $i$ at time $t$ in state $c(i,t)$:
If the local signal subject to Gaussian noise \textit{plus} the global signal subject to independent Gaussian noise exceeds a global threshold $a$,
then the cell state $c$ at the next time step $t+1$ is On, else Off.
[AND/OR]: If the local signal exceeds a local threshold AND/OR, then the global signal exceeds the global signal, $c(i,t+1)=$ On.
}
\label{fig:modelschematic}
\end{figure*}

\section{Model and observables}
\label{sec:model}

Our theoretical approach is based on a minimal model for the formation of a gene expression boundary in a single layer of cells.
The model focuses on the interplay between a morphogen gradient and local cell-cell signaling.
Consequently, it ignores all other processes occurring in developmental systems that might change the neighborhood relation of cells,
such as cell migration, proliferation, cell death, and cell shape changes.
Also, we treat the local coupling between cells completely on the level of information and do not consider additional physical mechanisms,
such as differential cell adhesion and differential mechanical tension \cite{Dahmann2011}. 
The role of the morphogen gradient is to activate the target gene in cells at positions where the morphogen level exceeds a threshold,
thereby creating a gene expression boundary.
By contrast, the local signaling provides a means for cells to exchange information about their gene expression state with their neighbors
and to use this information to modulate their response to the morphogen gradient.

Our model and the relevant observables are described in detail in the following two subsections.
On a conceptual level, it may be helpful to think of the model as a two-dimensional (2D) kinetic Ising model,
which has unusual couplings between neighboring spins and is subject to an inhomogeneous external magnetic field,
see the supplementary material for a detailed comparison \cite{Supplement}.
However, in the context of our study it is important that there are different sources of noise, not just a single thermal noise.
First, there is noise from the global morphogen signal.
In developmental systems, all processes involving the morphogen --- morphogen production, morphogen transport, morphogen uptake, and signaling
--- contribute to this noise.
For example, morphogen production is subject to stochastic variations in morphogen molecule synthesis and secretion \cite{Bollenbach2008, Raser2005},
morphogen transport by diffusion is a stochastic process itself and can moreover be hindered by barriers \cite{Restrepo2014}.
At the morphogen uptake stage, cell-cell variability in the number of receptors,
binding of molecules to receptors and receptor occupancy are additional sources of noise \cite{Colman-Lerner2005, Jaeger2008, Bollenbach2008}.
Finally, activation of the signaling pathways and gene regulation are also noisy processes \cite{Raser2005}.
These latter processes are also the cause for noise in cell-cell signaling.

\subsection*{Model}
We consider a square grid of $L\times L$ cells, see Fig.~\ref{fig:modelschematic}.
The state of a cell is reduced to either `On' or `Off', $c=+\frac{1}{2}\text{ or }-\frac{1}{2}$.
Along the axis of the gradient (index $i$) the boundary conditions of our grid are fixed to Off on the left ($i=0$)
and On on the right side of the grid ($i=L$).
In the perpendicular direction (index $j$) we apply periodic boundary conditions.
The cell state is updated according to a signal processing rule $\mathcal{L}$, which we also refer to as signal integration rule.
The state can change at discrete time steps.
The rule processes two different signals: the global signal at the cells position $(i,j)$, $\Gsij$,
representing the morphogen gradient,
and a local signal, $\Lsij$,
encoding the state of the cells within $\cij$'s neighborhood,
\begin{equation}
	\cij(t+1)= \mathcal{L}\left[\Gsij,\: \Lsij\right] \:.
\end{equation}
A Boolean logic is the common simplification of the biologically observed Hill type regulation,
as the sigmoidal form becomes a sharp threshold in the limit of large Hill coefficients \cite{Bolouri2002}.

\paragraph*{Global signal}
The stochastic global signal at a cell with index $(i,j)$ is given as
\begin{equation}
	\Gsij = m\: i + \GNij \: ,
\end{equation}
with $m$, $m\in [0,1]\subset\mathbb{R},$ the morphogen gradient slope at $i$ and $\GN$ additive Gaussian white noise
with mean zero and standard deviation $\Gstd$.

During embryogenesis, morphogen molecule concentration is commonly assumed to be exponentially decaying within the tissue.
Inspired by Ref.~\cite{Hillenbrand2016}, we interpret the logarithm of the molecule concentration as the actual signal (Weber-Fechner law),
resulting in a linear morphogen gradient signal.

\paragraph*{French Flag mechanism}
We refer to a signal processing rule that only depends on the global signal
and compares it to a global threshold $a$ as \textit{\Grad}\ rule $\mathcal{L}^\text{GRAD} \left[\Gs\right]$.
It is the analog of the French Flag mechanism \cite{Wolpert1969}.
More precisely
\begin{align}
\mathcal{L}^\text{GRAD} \left[\Gsij\right]&:=\Theta\left[\Gsij-a\right]-\frac{1}{2} \: ,
\end{align}
with $\Theta$ denoting the Heaviside step function with convention $\Theta\left[0\right]=0$.
The state of a cell at position $i$ at $t+1$ is $+\frac{1}{2}$ if the global signal exceeds the global threshold $a$ and $-\frac{1}{2}$ else.

\paragraph*{Local signal}
The signal processing rules with correction ability additionally make use of a local signal $\Ls$
that stems from nearest-neighbor cells communicating their state.
We conservatively assume that the central cell cannot sense from which neighbor the signal came from
and thus define $\Ls$ to be the sum of these signals
\begin{equation}
	\Lsij=\sum\limits_{(k,l)\:\in\: \text{neighbors}(i,j)}c_{kl}(t)\quad  +\LNij\: .
\end{equation}
`neighbors' refers to the upper, lower, left and right neighbor (von Neumann's neighborhood).
$\LN$ is chosen to be Gaussian white noise with a mean of zero and standard deviation $\Lstd$. Consequently, the noise realization is in $\mathbb{R}$, and with that the local signal.

\paragraph*{Correction mechanisms}

To implement a correction mechanism,
each cell needs to combine the two noisy signals $\Ls$ and $\Gs$.
It is by no means clear how this combination is optimally performed.
Straight forwardly, we can add up both signals
and compare the result to the global threshold $a$.
We will refer to this procedure as SUM rule $\mathcal{L}^\text{SUM}$,
\begin{equation}
	\mathcal{L}^\text{SUM}\left[\Lsij,\Gsij\right]= \Theta \left[ \Lsij+\Gsij -a\right] -\frac{1}{2} \: .
\end{equation}
Note that the contribution of the local signal to the full signal can take any value by rescaling $m$ and $a$ simultaneously.\\
Alternatively, both signals could be processed separately and the results combined by an AND or OR rule, denoted by $\mathcal{L}^\text{AND}$ and $\mathcal{L}^\text{OR}$, respectively.
A NOR or XOR rule is not expected to perform well in our setting,
as both signals are chosen to promote the On state, see also Appendix \ref{Supp:OtherLogicFunctions} for a more explicit discussion.
Processing the local signal separately requires an additional threshold,
$a^\text{AND}$ respectively $a^\text{OR}$,
\begin{align*}
	\mathcal{L}^\text{AND}\left[\Lsij,\Gsij\right]&= \Theta \left[ \Lsij-a^\text{AND}\right] \Theta\left[\Gsij -a\right] -\frac{1}{2}\:,\\
	\mathcal{L}^\text{OR}\left[\Lsij,\Gsij\right] &= \Theta \left[ \Lsij-a^\text{OR}\right]+\Theta \left[\Gsij -a\right]\\
	&\: -\Theta \left[ \Lsij-a^\text{OR}\right]\Theta \left[\Gsij -a\right] -\frac{1}{2} \: .
\end{align*}

If we want the AND and the OR rule to be able to produce a boundary from an arbitrary initial grid for all noise levels,
equivalently to the \Grad\ mechanism,
then we find that there is only one choice for the local thresholds,
i.e., $a^\text{AND}=-1$ and $a^\text{OR}=+1$.
For larger $a^\text{AND}$ values the AND rule cannot exit an initial all Off grid in the small noise scenario.
On the other hand, smaller $a^\text{AND}$ values increase the fuzziness as they reduce the correction ability of the AND rule,
as for $a^\text{AND}\ll-1$ we essentially have a \Grad\ rule.
A quantitative version of this argument and numeric confirmation are shown in Appendix \ref{Supp:alocal optimization}.
Equivalent reasoning holds true for the OR rule when starting from an all On grid.

For a choice of local thresholds such that $a^\text{AND}= - a^\text{OR}$ we observe in simulations
and can show,
a direct `particle-hole' correspondence between the AND and the OR rule,
resulting in
\begin{equation}\label{eq:ANDOReq}
	\langle \cij^\text{AND}\rangle \approx - \langle c_{\tilde{i}j}^\text{OR}\rangle , \quad\text{with}\quad
	\tilde{i}= 2\frac{a}{m}-i\: ,
\end{equation}
with the approximate relation becoming exact in the limit of an infinite grid.
The approximation generally works well for a boundary position that is distant from the grid boundary.
Essentially, this relation follows because the local signal is antisymmetric with respect to exchanging On for Off states
and linearity of the global signal, as made explicit in Appendix~\ref{Supp:AND-ORrelationship}.

These rules can also be motivated by common input functions to gene transcription \cite{Mayo2006, Buchler2003, Bolouri2002}:
Either both signals or their products within the signal processing pathway can occupy the same promoter,
a regulation scheme modeled by the SUM rule, motivated by Ref.~\cite{Kalir2004}.
Or there is a different promoter for each signal such that either both have to be occupied to switch on gene transcription,
representing the AND rule, or occupation of one is sufficient implying an OR rule.
The global threshold $a$ thereby corresponds to the binding affinity of the transcription factor stemming from the morphogen signal to its promoter
and the local thresholds to the binding affinity of the transcription factor related to the local direct neighbor signaling.
However, the model is not limited to regulation of a cell's transcriptional state by chemical signaling.
It also applies to bioelectrical signaling during embryogenesis for patterning processes involving a long ranged electric gradient,
where cell-cell signaling is performed via ion channels and gap junctions \cite{Levin2021}, for instance.
Or to mechanical cues as long as the cell neighborhoods are not altered.

\paragraph*{Transforming and reducing the parameter set}\label{par:parameterTrafos}
The parameters characterizing the model are the grid length $L$, morphogen gradient slope $m$, global threshold $a$,
and the standard deviations of the global and local noise $\Gstd$ and $\Lstd$.
In order to arrive at a description in more natural parameters,
we transform $m,\: a$ to $m,\:\frac{a}{m}$ as $\frac{a}{m}$ corresponds to the spatial position
where the morphogen signal equals its global threshold. Also, we transform the independent noises to a total noise and the relative contributions.
The total noise is defined as $\xi:=\LN+\GN$ with standard deviation $\sigma$ and local to total noise ratio $\alpha$ defined as
\begin{equation}
 \sigma^2 := \left(\Gstd\right)^2+\left(\Lstd\right)^2,\quad\text{and } \alpha := \frac{\left(\Lstd\right)^2}{\sigma^2} \: .
\end{equation}
If not stated otherwise,
then we set the local to total noise ratio $\alpha$ to $\alpha = \frac{2}{mL+2}$ as 2 is the maximal deterministic local signal
and $mL$ the maximal global signal.
\paragraph*{Simulation scheme}
For simplicity and computational efficiency,
we chose the dynamics to consist of synchronous updates of the complete grid at equidistant, discrete time steps.
We checked that the stationary state results qualitatively agree with a random update of all grid cells for the boundary position and fuzziness,
see the supplementary material \cite{Supplement}.

\subsection*{Observables}\label{sec:Observables}
\FloatBarrier
We are interested in a correctly positioned boundary between different cell types that is straight and sharp.
To make these notions quantitative,
we define the boundary position $\mathcal{B}(t)$ of a grid at time $t$ to be the average number of cells in Off state, per row,
\begin{equation}
	\mathcal{B}(t):=\frac{1}{L}\sum\limits_{i,j}\delta \left(\cij,-\frac{1}{2}\right) \:,
\end{equation}
with $\delta$ the Kronecker delta. This definition is related to the magnetization in Ising models and circumvents problems of other measures as discussed in the supplementary material \cite{Supplement}. 

The fuzziness $\mathcal{F}(t)$ of a grid at time $t$ is then defined as the number of sites in the wrong state
with respect to the boundary position, rounded to its closest integer, in percentage of the total number of cells,
\begin{equation}
	\mathcal{F}(t) := \frac{1}{L^2}\left(\sum\limits_{i<\mathcal{B}, j}\delta\left(\cij, \frac{1}{2}\right) + \sum\limits_{i>\mathcal{B}, j}\delta\left(\cij, -\frac{1}{2}\right) \right)\: .
\end{equation}
Note that this definition combines two notions characterizing the quality of a boundary,
its roughness and its softness.
Given a unique boundary line,
i.e., a grid configuration without holes,
the roughness quantifies the boundaries' deviation from a straight line.
In case of frequent holes,
the softness measures the width of the holey region that constitutes the boundary.
In the system presented here,
holes do occur, but are rare,
and thus we do not account for them separately.
The chosen definition of boundary fuzziness counts both types of boundary errors equivalently.
Exemplary grids visualizing boundary position and fuzziness can be found in the supplementary material \cite{Supplement}. 
The ensemble averages of the boundary position $\left\langle\mathcal{B}\right\rangle$
and the fuzziness $\left\langle\mathcal{F}\right\rangle$ are approximated by their time averages in the stationary state.

\FloatBarrier

\FloatBarrier
\section{Results}\label{sec:results}

For a boundary established by a global signal in the form of a gradient with slope $m$ and local signaling between neighboring cells,
we want to measure the dependence of the boundary position $\mathcal{B}$ and its fuzziness $\mathcal{F}$ on the total noise.
The total noise with standard deviation $\sigma$ sums independent Gaussian noise on the global and the local signal.

\subsection{Kinetics of approaching the stationary state}
We start our investigation of the correction mechanisms SUM, AND and OR
by studying the boundary position $\mathcal{B}(t)$ and fuzziness $\mathcal{F}(t)$ as a function of time
using a synchronous update scheme of the whole grid.
Toward that end,
we consider an arbitrary but fixed threshold $a$,
morphogen slope $m$,
and grid length $L$,
here chosen such that the boundary position of the \Grad\ mechanism is in the middle of the grid.

\begin{figure}
\centering
\includegraphics[width=1\linewidth]{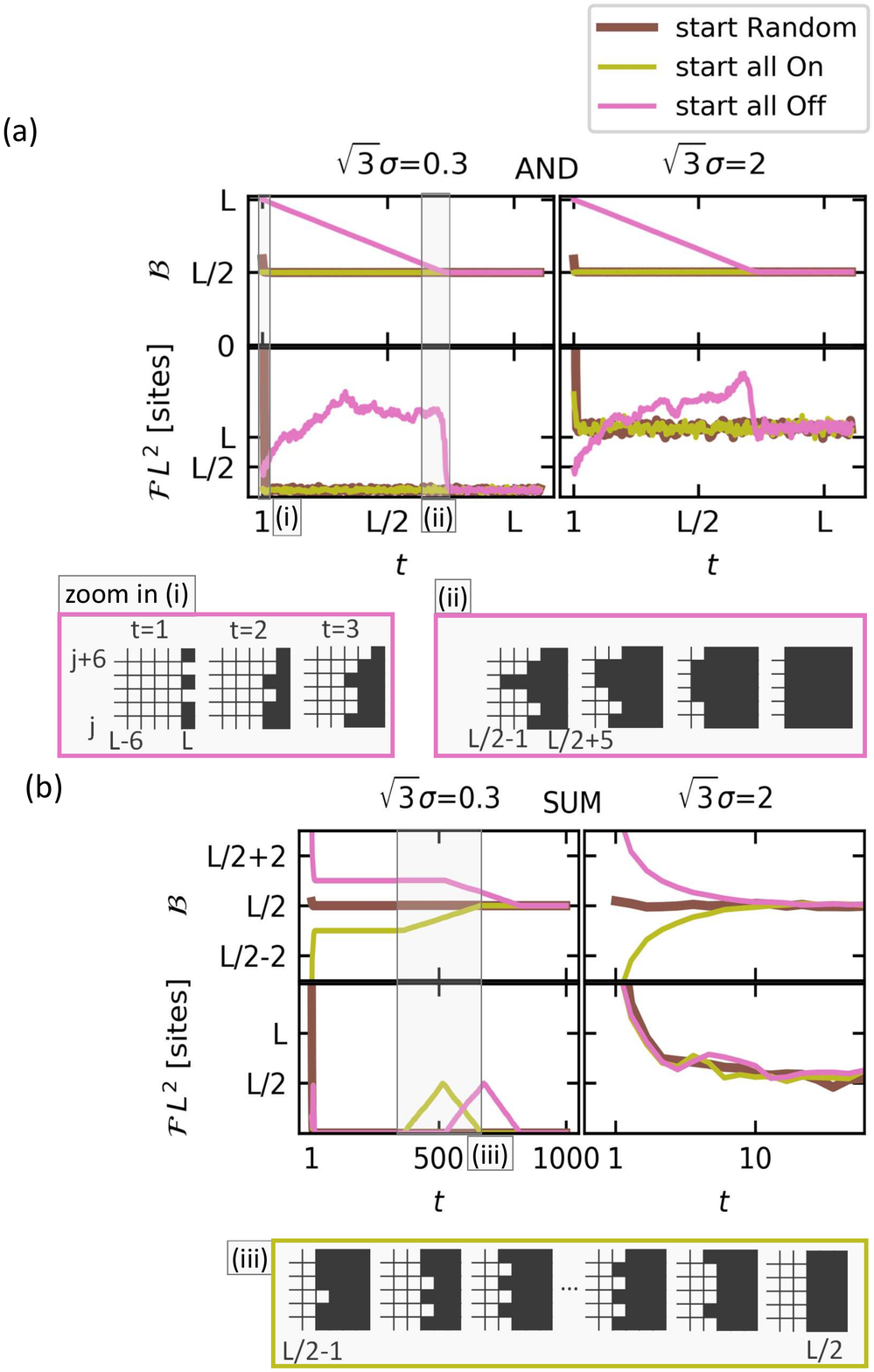}
\caption{Exemplary boundary position $\mathcal{B}$ and fuzziness $\mathcal{F}$ (in terms of number of wrong cells)
time traces starting from different initial conditions at $t=0$, depicted in different colors,
for the AND rule in the upper panel (a) and the SUM rule in the lower panel (b) subject to a small (first column)
and a large (second column) noise value.
Note that for the AND rule time is measured in terms of grid length.
The grid length $L=255$ is odd and $a=64.5$ and $m=0.5$ are chosen such that the stable boundary is in the grid center,
only $t\geq 1$ are shown for better visibility.
		Inserts (i)-(iii) provide a zoom in of 6$\times$6 sites to the grid, for few time steps. For the AND rule, starting from an all Off initial grid, zoom in (i) sketches the first three time steps, while zoom in (ii) provides a potential time trace of reaching the stable boundary position, which is $L/2$ here. For the SUM rule, zoom in (iii) sketches boundary destabilization and transition to a straight and sharp boundary. }
	\label{fig:timetraceplot}
\end{figure}

\paragraph*{AND and OR rule}\label{par:transitionAND-OR}

We characterize the evolution under the AND rule, plotted in Fig.~\ref{fig:timetraceplot}(a) for three different initial conditions:
a random initial grid,
a grid of all cells in state Off (`all Off'),
and a grid of all cells in state On (`all On').
The first column shows for an exemplary low noise level the boundary position in the first row
and the fuzziness in the second row
in dependence of time.

Starting from an all Off grid the boundary position $\mathcal{B}$ moves roughly one cell per time step
until it reaches its stationary value [see Fig.~\ref{fig:timetraceplot}(a), first row].
The fuzziness $\mathcal{F}$ increases until it drops sharply when the stationary boundary position has been reached.
Note that already for low noise the fuzziness time trace remains wiggly for all times, implying that the stationary boundary is fuzzy.
Remarkably, for larger noise the boundary position moves at the same rate.
Only the stationary boundary is more fuzzy
compared to the boundary in the low noise regime
[see right column of Fig.~\ref{fig:timetraceplot}(a)].
Consequently, the transition time to the initial condition independent, stationary, boundary position does not depend on the noise level.
We also observe that the transition time is on the order of magnitude of the stationary boundary position.
The last row of Fig.~\ref{fig:overviewplot} confirms this independence more generally.
\\
An intuitive picture, explained in more detail in the supplementary material \cite{Supplement}, 
for the dynamics is the following:
By definition, the AND rule only allows cells to switch on
if they have at least one On neighbor
and the gradient signal exceeds its threshold $a$.
The second condition is satisfied for all cells to the right of position $i=\floor{\frac{a}{m}}$ by the deterministic part of the gradient signal.
When starting from an all Off grid,
the first condition implies
that only cells at the right boundary can switch on
due to cells at $i=L+1$ being On (fixed grid-boundary condition choice).
Each boundary cell can move at most one cell forward per time step [see Figs. \ref{fig:timetraceplot}(i) and (ii)].
Quantitatively, a cell with exactly one On neighbor switches on with probability 1/2 independent of the total noise level.
A cell with more than one On neighbor switches on with a probability close to 1.
Note that our choice of grid-boundary conditions,
$c_{0,j}=0$ and $c_{L+1,j}=1$ for all rows $j$,
does not substantially simplify the patterning task for the rules.
The AND rule (and also the OR and SUM rule as we will see later) cannot just shift the sharp cell state boundary at $L$ to its stationary position.
Even if we had only one cell in On state at $i=L+1$ the rules could still establish a boundary at the center of the grid.
However, this choice of grid boundary would artificially destabilize the correct stationary pattern due to local signaling,
as the correct pattern requires that cells at $i=L$ are On.
\\
For the `all On' initial grid, convergence to the stationary boundary position occurs within the first time step, as the deterministic part of the neighborhood signal exceeds $a^\text{AND}=-1$ everywhere in the grid and the AND rule essentially reduces to the \Grad\ mechanism.
For a random initial grid, at $t=0$, the neighborhood signal exceeds $a^\text{AND}$ for about 3/4th of the cells in each column, thus convergence is fast, as well.

The OR rule by definition can only switch off one cell width at a time when starting from an initial grid of On cells.
Consequently, the time traces of the OR rule qualitatively correspond to the ones of the AND rule
with On-Off inverted initial conditions, see Fig.~\ref{fig:ANDOREquivalence} in the Appendix.
This was to be expected from the AND-OR relationship, Eq.~(\ref{eq:ANDOReq}).

\paragraph*{SUM rule}
Figure \ref{fig:timetraceplot}(b) shows evolution under the SUM rule and exhibits qualitatively different dynamics.
The right column shows the large noise regime.
We see that already within 20 time steps the different boundary position traces have converged.
We also note that the boundary is fuzzy in contrast to the low noise regime.
For a low noise level, as depicted in the left column of Fig.~\ref{fig:timetraceplot}(b), the dynamics is more complex.
While for all initial conditions,
the boundary quickly reaches a position close to its stationary value,
full convergence is very slow.
Thus, in contrast to the AND and OR rule the transition time strongly depends on the noise level.
Starting from any initial condition,
the boundary reaches a position close to its stationary value within few time steps.
Movement toward its final position happens
when noise induces a seed at the boundary linearly spreading until all cells within the same column have switched state.
As sketched in Fig.~\ref{fig:timetraceplot}(iii),
one seed induces a switch of both of its neighbors in the next time step and so forth
until the complete column of former boundary cells has switched state,
in case of an odd grid height.
Then the boundary remains straight until the next seed occurs.
In case of an even grid length,
the pattern with every second boundary cell switched on
corresponds to a metastable state, see Appendix \ref{Supp:EvenSystemLength}.
The waiting time distribution for the next seed to destabilize the metastable boundary is strongly noise level dependent.
In fact, the transition time until the stationary state at $\floor{\frac{a}{m}}$ is reached can be approximated by
\begin{equation}\label{eq:transitiontime}
	\mathcal{T}(\sigma) \approx \frac{12}{L}\exp\left(\frac{\sqrt{3}}{2\sigma^2}\right)
\end{equation}
for sufficiently small $\sigma$,
as shown in the supplementary material \cite{Supplement}, and plotted in Fig.~\ref{fig:overviewplot}.
This functional form clearly shows the nonlinear dependence of the transition time on the noise level.
As expected, the transition time is inversely proportional to the system length
as a seed is more likely the more sites in the column next to the boundary noise is acting on.

\FloatBarrier
\subsection{Characterizing the stationary state's dependence on the noise level}\label{Sec:StatState}

\begin{figure}[t]
\centering
\includegraphics{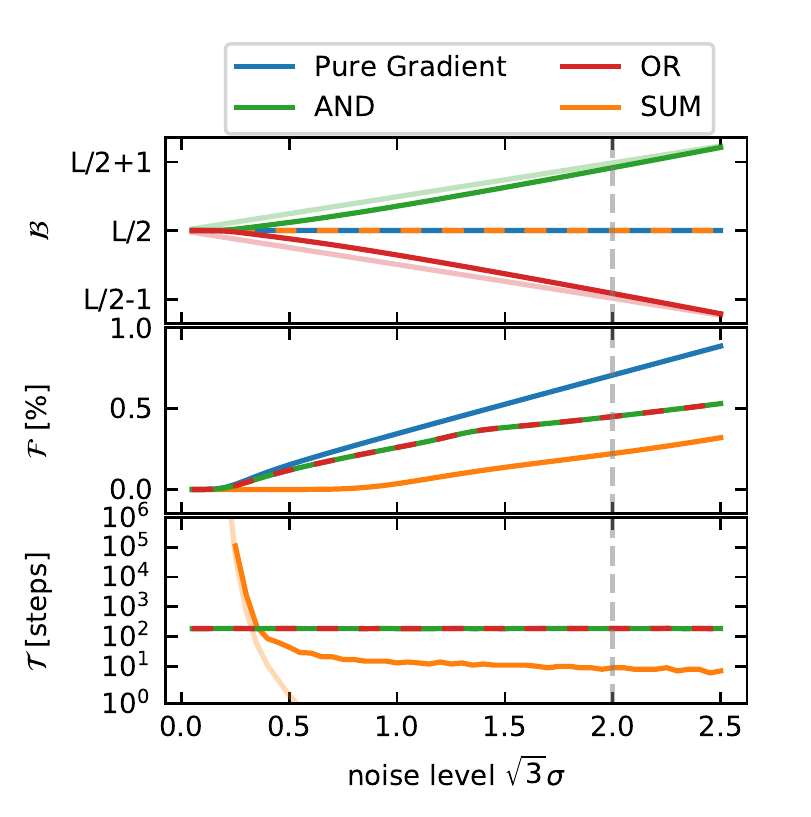}
\caption{
		Overview of different stationary state behavior of the \Grad , SUM, AND, and OR rule depending on noise level regime,
		exemplary for $L=256,\:a=64.5,\:m=0.5$.
		On the $x$-axis, we use \textit{noise level} $\nl$ instead of $\sigma$ for better intuition of the strength of the total noise:
		More than $90\%$ of all noise realizations are within the interval $\left[-\nl, +\nl\right]$.
		Also, a uniform distribution within $\left[-\nl, +\nl\right]$ has variance $\sigma^2$.
		In the first row, the time-averaged boundary position $\mathcal{B}$ for each rule is shown, where dashes indicate that the lines overlap.
		The light green and red lines show the analytical estimate of the boundary position for the AND and OR rule.
		In the second row, time-averaged fuzziness $\mathcal{F}$ in $\%$ of the total number of grid sites is plotted.
		The last row shows the transition time $\mathcal{T}$ until the initial condition independent state was reached,
		where a value of $10^6$ implies that it has not converged within the simulation time. The light yellow line depicts the $\mathcal{T}$ approximation for the SUM rule for small noise levels, see Eq.~(\ref{eq:transitiontime}).
	}
	\label{fig:overviewplot}
\end{figure}

Figure~\ref{fig:overviewplot} shows the characteristic stationary state properties of the three correction mechanisms SUM, AND, and OR
as a function of the total noises' standard deviation $\sigma$ for an exemplary threshold $a$ and morphogen signal slope $m$ choice.
For direct comparison,
the results without correction mechanism (\Grad) are plotted in blue.
Each row shows a different observable ---
the time-averaged boundary position $\mathcal{B}$, fuzziness $\mathcal{F}$, and,
in the last column, the transition time $\mathcal{T}$.
For simplicity, we here use the \textit{maximal} number of time steps until reaching the stationary state from an all On and an all Off grid
as a measure of $\mathcal{T}$.
We will discuss the different phenomenologies starting from low noise levels and ending with large noise levels.

\paragraph*{Small noise levels}\label{sec:FuzzFreeSUM}

In the last row of Fig. \ref{fig:overviewplot} we observe that
the SUM rule results have not converged to stationary state
within the simulation time of $10^6$ time steps
in the regime of very low noise levels.
This is expected from the previous transition time discussion.
Loosely speaking, noise is needed to forget the initial grid state.
In contrast, the AND and OR rule's transition time $\mathcal{T}$ scales linearly with the boundary position,
irrespective of the noise level as discussed in Sec.~\ref{par:transitionAND-OR}.
The transition time from a \textit{particular} initial condition, e.g., `all On', can be significantly smaller, see Fig.~\ref{fig:timetraceplot}.
Turning to the second row of Fig.~\ref{fig:overviewplot},
we note that the boundary fuzziness for the SUM rule is remarkably close to zero
for small yet sufficiently large noise levels to allow for convergence of the SUM rule pattern.
An effectively nonfuzzy regime is not observed for other rules.
In the first row, we observe that the SUM's boundary position agrees well with the \Grad\ rules' for zero noise,
while the AND and OR rules' boundary position slightly deviate up to a cell width.
Indeed, we can derive analytic approximations for the stationary boundary positions, as outlined below, and refer to Appendix \ref{Supp:AnalyticsStationaryBP} for further details. For the SUM rule, the stationary boundary position is given by
\begin{equation}
	i_c^\text{SUM} = \floor{\frac{a}{m}} = i_c^\text{Grad} \: ,
\end{equation}
with $\floor{\cdot}$ denoting the floor operator.
Consequently, the stationary boundary will scale with system size in the same way as for the \Grad\ rule for zero noise.
The stationary boundary position of the AND rule includes an additional term linear in the noise level,
\begin{align}\label{eq: AND_BP}
	i_c^\text{AND}\approx \floor{\frac{a}{m}}+0.25\frac{\nl}{m}\sqrt{1-\frac{2}{2+mL}} \: .
\end{align}
We see that $i_c^\text{AND}\approx i_c^\text{Grad} +\frac{\nl}{4m}$ for $mL\gg 2$
which agrees well with simulation results shown in Fig. \ref{fig:overviewplot}. \\
For the OR rule, it follows
\begin{equation}\label{eq: OR_BP}
	i_c^\text{OR} \approx \floor{\frac{a}{m}}-0.25\frac{\nl}{m}\sqrt{1-\frac{2}{2+mL}} \: ,
\end{equation}
by the AND-OR equivalence established in Eq. \ref{eq:ANDOReq}.
To derive those expressions for the stationary boundary positions, we started from the following observation:
The probability for a defect to disturb the boundary at $i_c$ needs to be smaller than the one for a potential boundary one cell width to its left,
at $i_{c-1}$, or to its right, at $i_{c+1}$.
Those disturbances can be either an Off cell right of the boundary (Off-in-On defect) or an On cell left of the boundary (On-in-Off defect).
For a sketch of an Off-in-On defect, see the first grid of Fig. \ref{fig:timetraceplot}(iii).
Consequently, two conditions need to be satisfied such that the stationary boundary position is at cell index $i = i_c$.
To the left, the probability to destabilize a boundary at $i_c-1$ by an Off-in-On defect has to be smaller or equal
to the probability destabilizing a boundary at $i_c$ by an On-in-Off defect.
To the right, the probability to destabilize a boundary at $i_c$ by an On-in-Off defect has
to be smaller or equal to the probability destabilizing a boundary at $i_c+1$ by an Off-in-On defect.
Solving these inequalities with probabilities approximated for the different rules yields the stationary boundary position;
see Appendix \ref{Supp:AnalyticsStationaryBP}.
From this calculation we can also see
that the probabilities for an On-in-Off defect and an Off-in-On defect at $i_c$ depend only on $m$ and $\frac{a}{m}-\floor{\frac{a}{m}}$.
These findings suggest that the fuzziness in stationary state only depends on the deviation of $\frac{a}{m}$ to the next integer value.
This is also confirmed by numeric results.
Intuitively, the morphogen changes at the same rate everywhere in the system
and the local interaction is independent of the position per definition.

\paragraph{Intermediate noise levels}
For larger noise levels the AND and the OR rule qualitatively exhibit the same behavior as for low noise, in contrast to the SUM rule.
In the second row of Fig.~\ref{fig:overviewplot} we observe a rapid increase in fuzziness for the SUM rule.
Time-averaged fuzziness seems to arise from alternating between time intervals of a straight,
sharp boundary and time intervals with a disturbance,
seeded by a single defect cell,
that grows and shrinks for some time before it decays.
The probability for a seed is highly nonlinearly, but smoothly, increasing with noise level \cite{Supplement}.

\paragraph{Large noise levels}
In the regime of large noise
we see in the second row in Fig. \ref{fig:overviewplot} that the SUM rule yields a less fuzzy boundary than the AND and OR rule,
which behave similarly.
Indeed, all correction mechanisms outperform the \Grad\ rule.
We will show that this result is robust for all $a$, $m$ parameter combinations determining the three rules in Sec.~\ref{sec:FuzzFrac},
for all grid lengths (Sec.~\ref{sec:Scaling}) and for a surprisingly large range of local to total noise ratios $\alpha$
(Sec.~\ref{sec:VaryAlpha}).

The boundary position $\mathcal{B}$ resulting from the SUM rule agrees well with the boundary position from the \Grad\
as analytically deduced in Appendix~\ref{Supp:AnalyticsStationaryBP}.
Although the boundary positions from the AND and OR rules deviate linearly with the standard deviation of the total noise $\sigma$,
we will show in Sec.~\ref{sec:Scaling} that both nevertheless scale linearly with system size.

\subsection{All rules conserve scaling of the boundary position with system size}\label{sec:Scaling}

\begin{figure}
	\centering
	\includegraphics{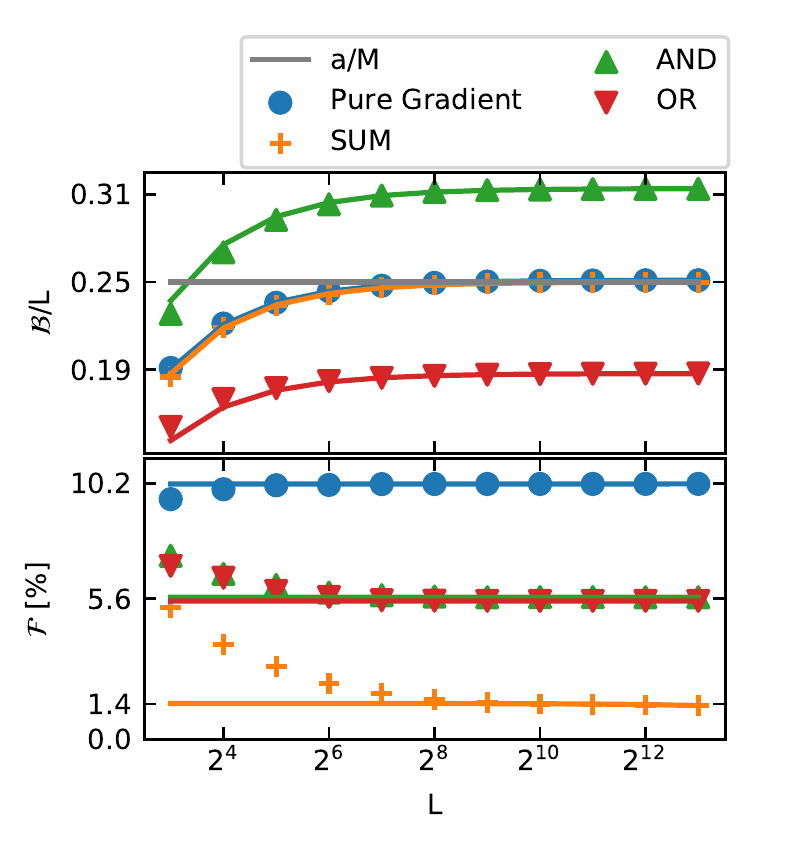}
	\caption{
		Upper panel: Relative time-averaged boundary position $\mathcal{B}$ for all three rules and \Grad\
		for increasing grid lengths $L$
		for a fixed noise level of $\nl=2$, $a=2,\: mL=8,$ and all Off initial grid.
		Bottom panel: Time-averaged boundary fuzziness $\mathcal{F}$ in percentage of the number of cells in the grid.
		For sufficiently large grids the relative fuzziness converges to a constant value that is smallest for the SUM rule,
		larger for the AND and OR rule and highest for the \Grad\ rule.
	}
	\label{fig:scaling}
\end{figure}

In embryogenesis, proportions commonly remain the same irrespective of different embryo or compartment sizes
(see \cite{Inomata2017a} for a review).
The \Grad\ mechanism also exhibits this scaling behavior,
provided that the maximal morphogen concentration and the threshold remain constant.
In our model these conserved proportions translate to a fixed fraction of On to Off cells within grids of different size
for a fixed maximal morphogen signal $mL$.
In Fig. \ref{fig:overviewplot} we have observed that the stable boundary position formed by the AND and OR rule
deviates from the position established by the \Grad\ rule.
Thus we need to investigate if also the AND and OR rule ensure this property, just with a different fraction.
For an exemplary parameter set, $a=2,\: mL=8$ at a large noise level $\nl=2$,
we can see in Fig. \ref{fig:scaling} that this is indeed the case.
The reason that AND (OR) rules' boundary position tends to larger (smaller) $\mathcal{B}$ values is that it discourages (encourages) On cells.
As the slope $m$ becomes smaller and smaller ($mL$ fixed) the regime around the boundary in which this effect plays a role
increases linearly with system length.
This leads to a constant boundary position $\mathcal{B}$ over grid length $L$ ratio.
The deviation of this value from the \Grad\ rules' result can be approximated by Eq. (\ref{eq: AND_BP}) and Eq. (\ref{eq: OR_BP}), respectively.
For small grid sizes,
we see that the relative boundary position of any rule has not yet converged to its large grid limit.
The reason is of technical nature: Our parameter choices imply that defect cells at the left of the boundary are more common than at its right,
see Appendix \ref{Supp:ScalingConvergence} for a more detailed argument.

Figure \ref{fig:scaling} shows that the fraction of sites in the wrong state converges to a stable value for large system lengths.
Consequently, the improved boundary sharpness is not only a finite grid size effect.
The observation that the SUM rule performs best, the AND and OR rule not as well,
but better than \Grad , also holds true for all tested grid lengths.

\subsection{Systematic exploration of the parameter space}\label{sec:FuzzFrac}

\begin{figure}
\includegraphics[width=0.9\linewidth]{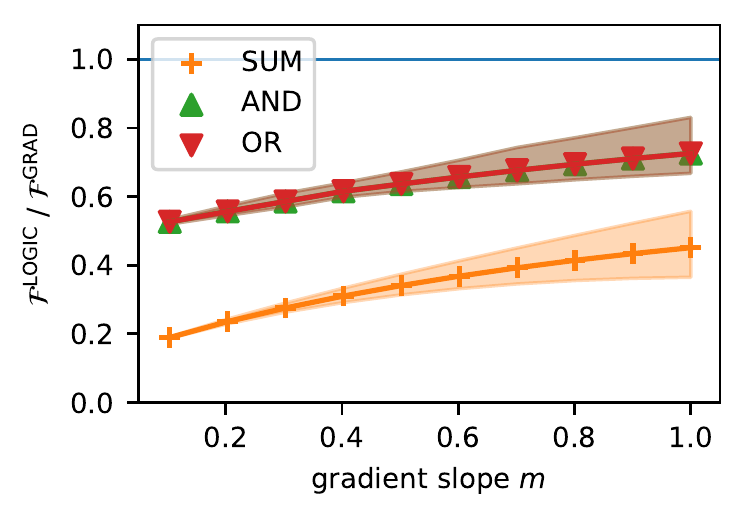}
\caption{For a fixed noise level $\nl=2$ and grid length $L=256$ the fuzziness $\mathcal{F}$ fraction of the boundary formed by a correction rule
(SUM, AND, OR) and the boundary fuzziness due to \Grad\ is shown in dependence of the gradient slope $m$.
Single markers correspond to the average fuzziness fraction over all $a$ values and are interpolated by a solid line.
The shaded area is restricted by the interpolation of the minimal (lower edge) and maximal (upper edge) fuzziness fraction
with respect to all threshold $a$ values. }
\label{fig:fuzzfrac}
\end{figure}

We want to know
how much smoother the boundary established by the correction mechanisms is
compared to the boundary established by the \Grad\ rule.
Until now, we have shown results for isolated points in the parameter space.
Now we want to study the whole parameter space spanned by threshold $a$ and morphogen signal slope $m$.
It is equivalent to varying $m$ and $\frac{a}{m}$ from $\floor{\frac{a}{m}}$ to $\ceil{\frac{a}{m}}$ as discussed in Sec.~\ref{Sec:StatState}.\\
Let us fix a high noise level, $\nl=2$.
We measure the boundary smoothing capability of a correction rule in terms of the ratio of the boundary fuzziness
resulting from a correction mechanism to the boundary fuzziness caused by the \Grad\ rule,
$\mathcal{F}^\text{LOGIC}(\nl=2)/\mathcal{F}^\text{GRAD}(\nl=2)$.
In order to show the full range of fuzziness ratios caused by varying $a$ and $m$ in Fig.~\ref{fig:fuzzfrac},
we choose for each $m$ to display the ratios' minimal and maximal value for any $a$
(interpolated by the lower and upper edge of the shaded area), as well as the ratios' average over $a$ (data points interpolated by sold line).

We observe that $\mathcal{F}$ ratios are below one and that the SUM ratio is smallest for all morphogen slopes $m$.
This implies that all three rules perform better for every threshold $a$ and gradient slope $m$
than the \Grad\ rule with respect to fuzziness reduction.
The performance gap between the \Grad\ and the correction rules is smaller for larger morphogen slopes,
and thus the most conservative choice for $m$ is $m=1.0$.
Intuitively, this is because the steepest gradient provides the largest signal differences between neighboring cells in $i$ direction.

\subsection{Variation of local to total noise ratio}\label{sec:VaryAlpha}
\begin{figure}
	\centering
	\includegraphics{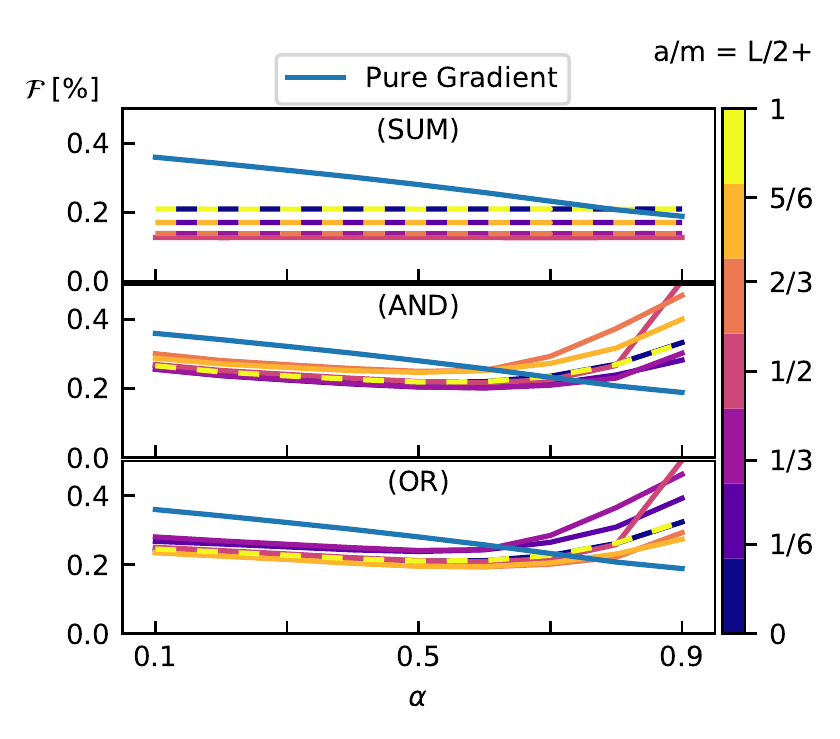}
	\caption{
		Fuzziness $\mathcal{F}$ at $\nl=2$ in dependence of the local to total noise variance,
		$\alpha$ for $m=1.0$, one row for each correction mechanism.
		The blue line depicts the \Grad\ performance,
		while the differently colored lines correspond to the correction mechanism result for different $a$/$m$ combinations, $L=256$.
	}
	\label{fig:varyalpha}
\end{figure}
The magnitude and ratio of local to global noise representing a variety of different processes as mentioned in the Introduction is not known.
Still, for a fixed maximum morphogen signal $mL$,
we have only considered one particular ratio of the local noise $\LN$ to total noise $\LN+\GN$ ratio $\alpha$,
$\alpha = \frac{(\Lstd)^2}{(\Lstd)^2+(\Gstd)^2}=\frac{2}{2+mL}$ so far.
Intuitively, the \Grad\ rule should perform better than the correction mechanisms for large $\alpha$.
By definition, the \Grad\ rule only experiences noise on the global signal,
which approaches zero for the local to total noise ratio $\alpha$ approaching 1.
The correction mechanisms, however, process an additional highly error-prone signal, the local signal.
To test this intuition, we vary the local to total noise ratio $\alpha$ between 0.1 and 0.9, see Fig. \ref{fig:varyalpha}
for all different $\frac{a}{m}$ combinations with  $m=1.0$, depicted in different colors.
The maximal slope $m=1.0$ is chosen to ensure the best relative \Grad\ performance, as discussed in Sec. \ref{sec:FuzzFrac}.

Per definition, the SUM rule (first row) is independent of $\alpha$, as it adds up the Gaussian distributed local $\LN\sim \mathcal{N}(0,(\sigma^L)^2)$ and global noise $\GN\sim \mathcal{N}(0,(\sigma^G)^2)$, yielding $\LN+ \GN \sim \mathcal{N}(0,\sigma^2)$.
The AND and OR rule perform worse relative to the SUM rule the larger $\alpha$ is.
Surprisingly, the $\alpha$ value, where \Grad\ (blue line) performs equally well than the correction mechanisms
(other colors) is well beyond one half.
This implies that a correction mechanism can sharpen the boundary, even if it is subject to more than twice as much noise than the \Grad\ rule.
%

\section{Discussion}\label{sec: summary}
\FloatBarrier

Precise boundary formation is a remarkable phenomenon in developing systems
and aspiring goal in synthetic systems.
Here our objective was not to study any specific system,
but to systematically explore how different logical couplings of cell-cell communication with a gradient signal can aid boundary formation.

In a minimal model of boundary formation consisting of cells that are either signaling (On) or inactive (Off),
we studied three different correction mechanisms using this nearest-neighbor interaction
in addition to a global (grid spanning) signal gradient.
Those three rules either sum both signals (SUM) and subsequently compare the result to a global threshold $a$
or compare each signal to a local and global threshold separately (AND, OR).
Consequently, for a signal processing cell to switch to or maintain an On state,
in case of the SUM rule the sum of both signals has to exceed the global threshold.
The AND rule requires both signals to exceed their respective thresholds independently,
while the OR rule requires only one signal to exceed its threshold.
We examined which rule performs best in which regime of total noise, consisting of additive Gaussian noise on the local and the global signal.
As motivated in the Introduction, performance is measured in terms of
(i) reduction in boundary fuzziness,
(ii) short transition time to stationary boundary position,
(iii) position tuneable by threshold, and
(iv) scaling with system size.

We found that
(i) the SUM rule achieves the strongest fuzziness reduction,
while the AND and OR rule yield comparably less reduction.
Only if the noise on the global signal is much smaller than the noise on the local signal,
the \Grad\ ensures a sharper separation of cells with different gene expression states.
(ii) However, transition to stationary state takes more time for any correction mechanism than for the \Grad\ rule,
which establishes the stationary boundary within one time step.
The SUM rule generates a boundary position that deviates by few sites within one time step for sufficiently steep morphogen slopes.
Exact convergence strongly depends on the noise level --- the smaller, the slower.
Qualitatively different,
convergence of the AND and OR rule does not depend on the noise level,
but on the initial state of the grid.
In the best case it happens within one time step, in the worst case in order of grid length $L$ time steps.
A short transition time is desirable in development,
as boundary cells often act at organizing cells for the next patterning process \cite{Dahmann2011}.
Also, fast morphogenesis is favorable to protect against predators.
(iii) The boundary position can be tuned by changing the global threshold value $a$ in a similar manner than for the \Grad\ mechanism.
This programability is of biological significance,
as the threshold $a$ was motivated by the binding affinity of the signals' transcription factor to the promoter of the gene, that is switched on.
Possibly, variations of same theme among related species,
such as a stripe that differs in width,
can be explained as variations of the binding affinity.
(iv) For fixed threshold value $a$ and morphogen signal slope $m$ the SUM rule yields the same boundary position as the \Grad,
while the AND and OR rules' boundary positions deviate by a small amount.
Nevertheless, the boundary position established by any rule scales linearly with system size.
The scaling property ensures compatibility with the observation that embryos of the same species differ (slightly) in size,
but pattern ratios are often conserved \cite{Inomata2017a}.

What is the underlying reason that the SUM rule outperforms the AND and OR rule in terms of fuzziness reduction?
We believe this is because the SUM rule first averages the involved noises allowing them to cancel,
before applying the nonlinearity in the form of comparing to a threshold.
The deterministic signal is summed, while the Gaussian noise's standard deviations only add in quadrature.
For the AND and OR rule it is the other way round:
They threshold each signal separately before combining the pieces of information.
However, there are biological problems where this signal processing scheme
--- first thresholding, and then combining ---
is actually optimal,
for example in case of a rare signal.
Rod cells in the retina specialized to detecting very dim light
first threshold before transmitting information to their common bipolar cell,
which consecutively sums those digitized signals \cite{Field2002a}.
If the bipolar cell would first sum the signal of its numerous rod cells,
then the simultaneously summed up noise would make it likely for the bipolar cells to confuse total darkness (no photon)
with dim light (three or more photons).
Consistent with the system presented here,
it has been found that only those rod cells specialized for dim light detection process signal by thresholding before summing \cite{Bialek2012}.

In a synthetic setup it is feasible to experimentally test our predictions as all necessary components have already been designed.
On the one hand, tuning of local interactions has been successfully realized in multicellular systems,
for example, AND-like, OR-like and in between, graded, SUM-like, regulatory behavior in yeast \cite{Khalil2019}.
A remarkably customizable signal processing scheme in the form of a synthetic Notch pathway is presented in Ref.~\cite{Toda2019},
and also reviewed in Ref.~\cite{Elowitz2021} among other protein based synthetic circuits in eukaryotic cells.
On the other hand, tuneable processing of synthetic gradients has been demonstrated,
such as toggle switch processing of a signal gradient in \textit{Escherichia coli} \cite{Barbier2020}.
A setup close to developmental biology was constructed within \textit{Drosophila} wing primordia \cite{Stapornwongkul2020}.
A combination of a synthetic gradient and a signal processing pathway that can in principle be customized to the rules discussed in this paper
is the synthetic GFP morphogen that regulates target gene expression by a synNotch circuit \cite{Toda2020}.

Remarkably, the study presented in Ref.~\cite{Stapornwongkul2020} raises the question of
which additional mechanisms are required for sharp boundary formation as the explored setups did not give rise to sharp boundaries \cite{Barkai2020}.
The toolboxes listed above might be able to test whether the three different rules studied in this paper
are candidates for the actual boundary correction necessary to ensure the exact results observed in nature.
Also, insights about the potential of local signaling as a correction mechanism might find applications in synthetic biology more generally,
as boundary and stripe formation are fundamental tasks in patterning and morphogenesis.
To better meet such applications, the presented model can be trivially extended to stripe generation by introducing extra thresholds accordingly.
It can be further generalized or modified by including more complex rules or smoothing the hard threshold via Hill-type functions,
by considering irregular and dynamic cell grids, or by including additional mechanisms such as cell sorting.


\begin{acknowledgments}
We thank Leo Stenzel, Gasper Tkacik, Nicolas Gompel and Hamid Seyed-Allaei for insightful discussions,
as well as Cesar Lopez-Pastrana, Isabella Graf and Johannes Harth-Kitzerow for critical reading of the manuscript.
This research was supported by the German Research Foundation via the collaborative research center SFB1032 via U.G.
MB was supported by a DFG fellowship through the Graduate School of Quantitative Biosciences Munich (QBM).
\end{acknowledgments}
\cleardoublepage
\appendix

\newcommand{\p}[1]{\left(#1 \right)}
\renewcommand{\d}{\text{d}}
\renewcommand{\emph}{\textit}
\FloatBarrier
\cleardoublepage

\FloatBarrier
\section{Other rule functions}\label{Supp:OtherLogicFunctions}
SUM, AND, and OR rules are not the only options to process two signals.
Others are XOR and PROD, but we can argue that they are not suited for the boundary formation problem as we modeled it. \\
Let us start from an all Off grid with an XOR logic. In the next time step without noise it would form the correct boundary. In the consecutive update step, all On cells except those at the boundary would switch Off though, as each is subject to a local neighbor signal greater than any (sensible) local threshold value. Consequently the boundary would not be stable.\\
The product rule PROD in the presence of noise reads\\
\begin{tabular}{l l}
	if & (global signal(i) +$\xi^G$)$\cdot$ (local signal(i,t) + $\xi^L$) $>\: a^2$:\\
	&\quad cell(i, t+1) = On
\end{tabular}
which implies that the noise would be multiplied by the signal.
Consequently, we expect this rule to perform poorly in the presence of sufficiently large noise.

\section{$a_\text{local}$ optimization}\label{Supp:alocal optimization}

We want the AND and OR rules to be able to produce a boundary
from an arbitrary initial grid
for all noise levels,
equivalently to the \Grad\ mechanism.
Here we show that the initial grids all Off and all On are sufficient
to fix the additional local thresholds $a^\text{AND}$ and $a^\text{OR}$.

Let us consider an all Off initial grid.
At the right border, $i=L$, the global signal exceeds the global threshold $a$
(otherwise the \Grad\ rule could not form a nontrivial boundary either).
For the AND rule to exit the initial condition,
we need $a^\text{AND}$ to be smaller or equal than the local signal $\Lsij=-1+\LNij$.
Thus, we need $a^\text{AND}\leq -1$.
Similarly, for an all On initial grid it follows that $a_l^\text{OR}\geq 1$.
The second condition to determine the optimal local threshold comes from demanding that it stabilizes a straight boundary.
To this end, consider a straight boundary, implying that the global signal is close to $a$, but with one On cell left of the boundary.
For the AND rule, we want $a^\text{AND}\geq -1$ in order to switch Off the defect cell.
Taken together this suggests choosing $a^\text{AND}= -1$. The opposite scenario, one Off cell right of the boundary, yields $a^\text{AND}<1$,
which is well satisfied by our choice.
Equivalent reasoning yields $a^\text{OR}=1$.

\begin{figure}
	\centering
	\includegraphics[width=1\linewidth]{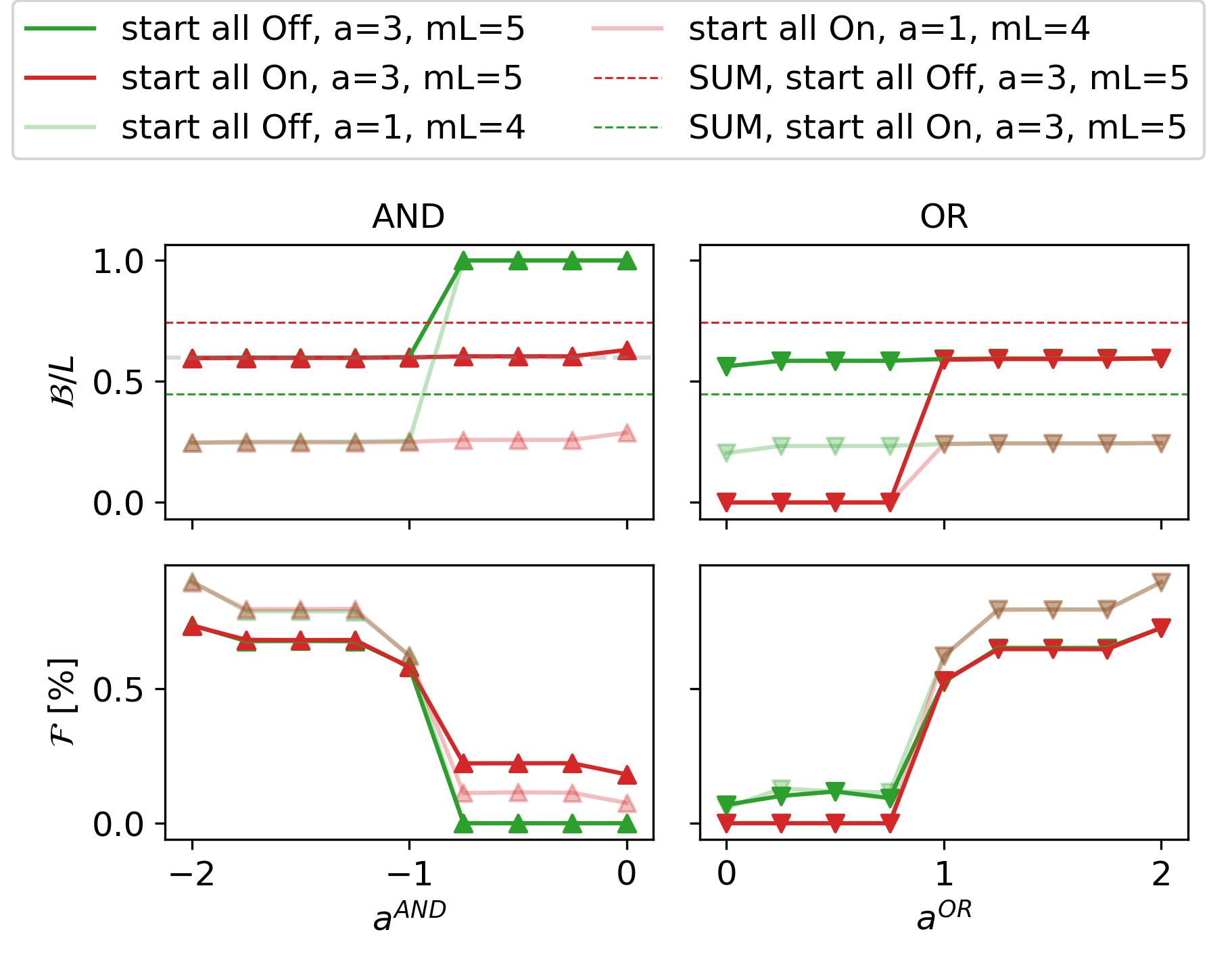}
	\caption{Left column: Relative boundary position $\mathcal{B}/L$ and fuzziness $\mathcal{F}$ with respect to the local threshold $a^\text{AND}$
	for two different $a$ and $m$ combinations, one plotted full saturation, one light.
	Green (red) triangles show results for an initial all Off (all On) grid, interpolated by solid lines.
	The dashed gray line shows the \Grad\ $\mathcal{B}/L$.
	The first column shows that the all Off initial condition can only be exited for $a^\text{AND}\leq -1$.
	The second row shows the fuzziness decrease with increasing $a^\text{AND}$. Right column:
	Analogous results for the OR rule. A small noise level of $0.1=\eta=\sqrt{3}\sigma$ is used.}
	\label{fig:alocaleta0dot1}
	\centering
	\includegraphics[width=1\linewidth]{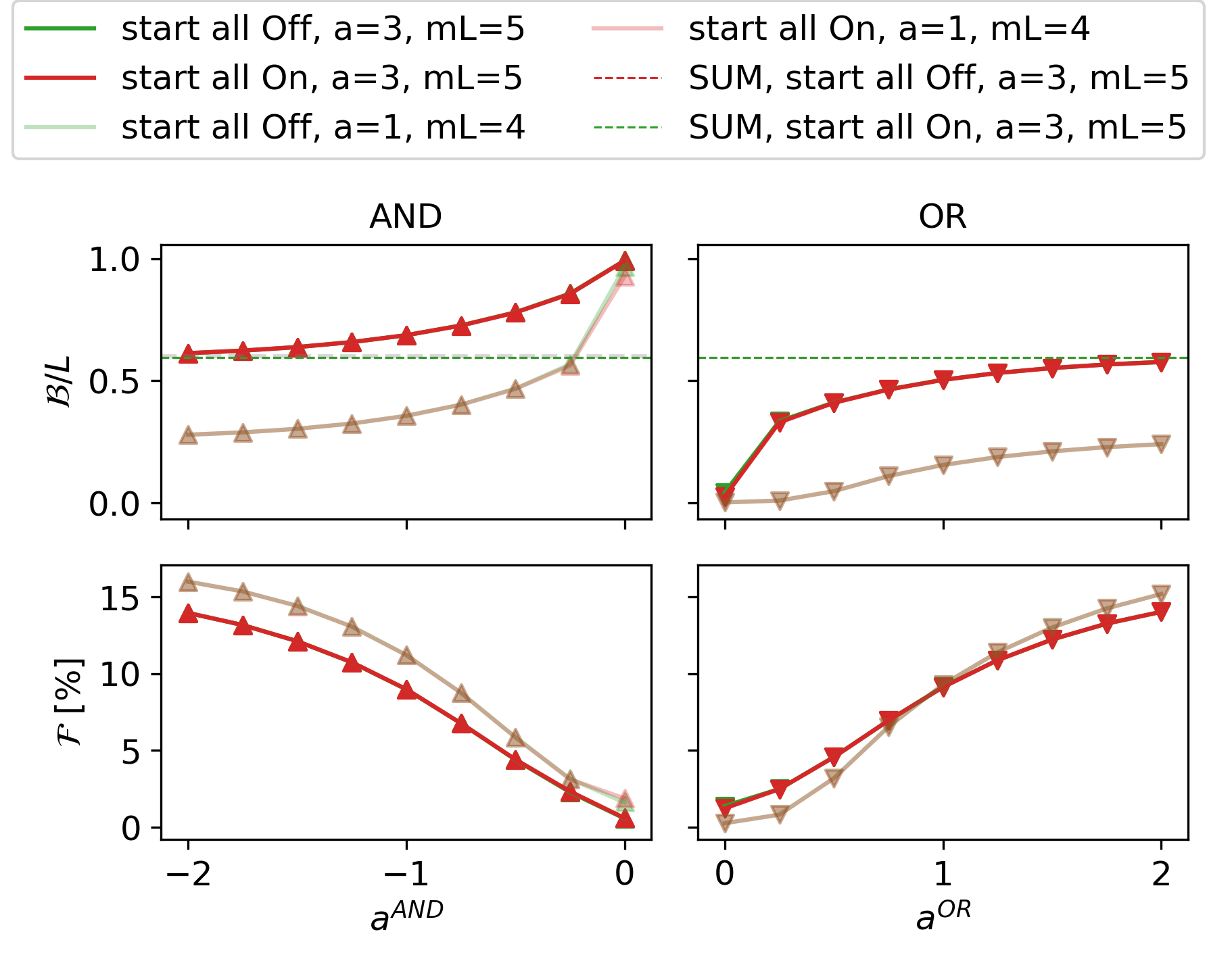}
	\caption{Same quantities as in Fig.~\ref{fig:alocaleta0dot1} but for a large noise level.
	Here, both initial conditions yield the same boundary position independent of $a^\text{AND}$ choice,
	but the boundary position deviates strongly from the \Grad\ value.}
	\label{fig:alocaleta2}
\end{figure}

We confirmed these analytic arguments numerically for an exemplary small and large noise level, see Fig. \ref{fig:alocaleta0dot1} and Fig. \ref{fig:alocaleta2}.
We observe that the fuzziness decreases with increasing $a^\text{AND}$ values.
In the small noise example, $a^\text{AND}=-1$ is the largest $a^\text{AND}$ value
such that the AND rule forms a nontrivial boundary (i.e., $\frac{\mathcal{B}}{L}\neq 1$ or 0)
independent of starting from an all On grid (red line) or an all Off grid (green line).
If we drop the initial grid independence condition,
e.g., if it suffices that the AND rule only patterns when starting from an all On initial grid,
then the optimal choice of $a^\text{AND}$ would depend on the magnitude of deviation of boundary position from the one generated by \Grad\ rule
(dashed gray line) we are willing to accept.
This is deviation is particularly pronounced for large noise as shown in Fig.~\ref{fig:alocaleta2}.
Results for the OR rule are depicted in the right column and findings are analogous.

\FloatBarrier
\section{Relationship of AND and OR rule for $a^\text{AND}= -a^\text{OR}$}\label{Supp:AND-ORrelationship}
\begin{figure}
	\centering
	\includegraphics[width=0.9\linewidth]{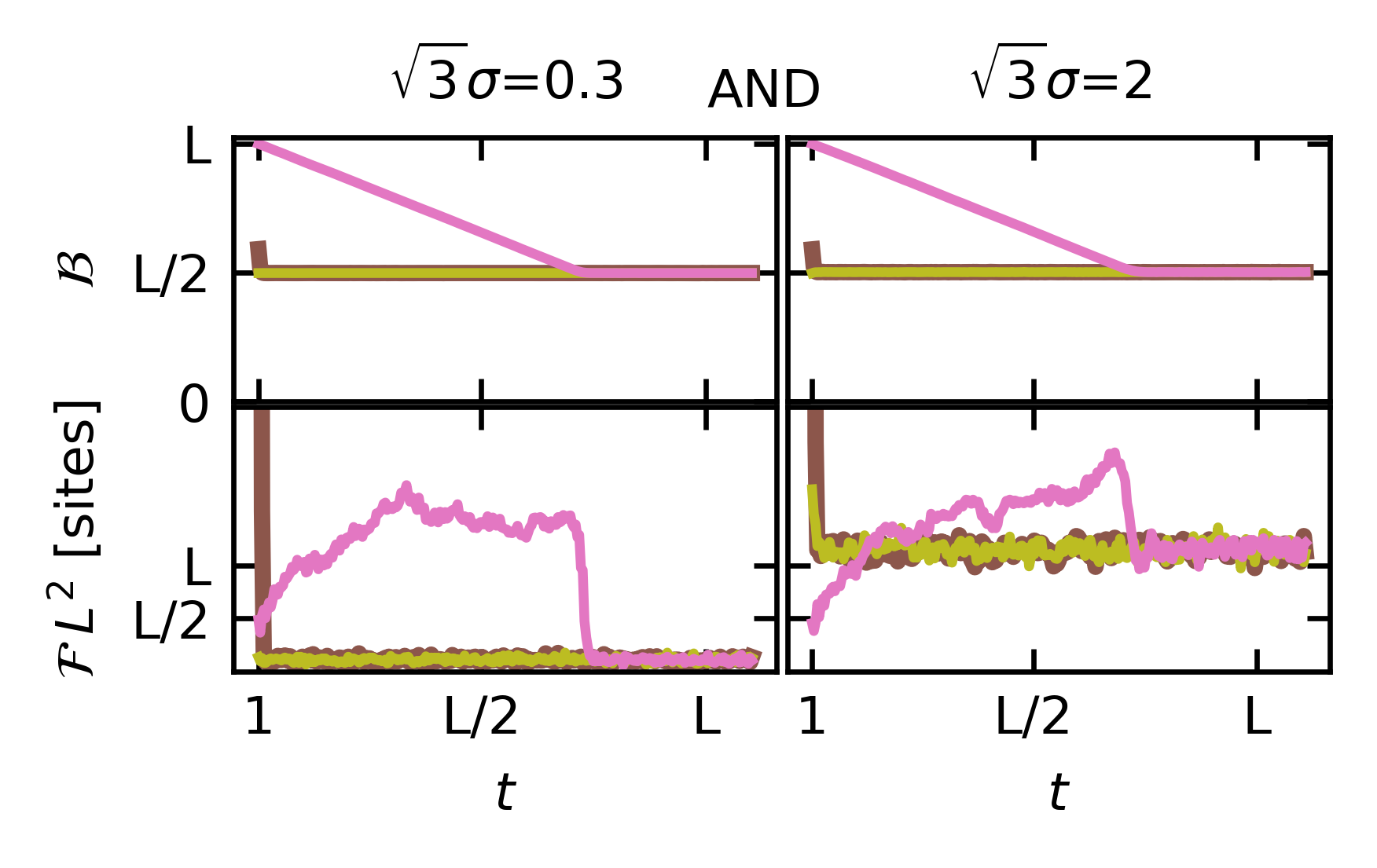}
	\includegraphics[width=0.9\linewidth]{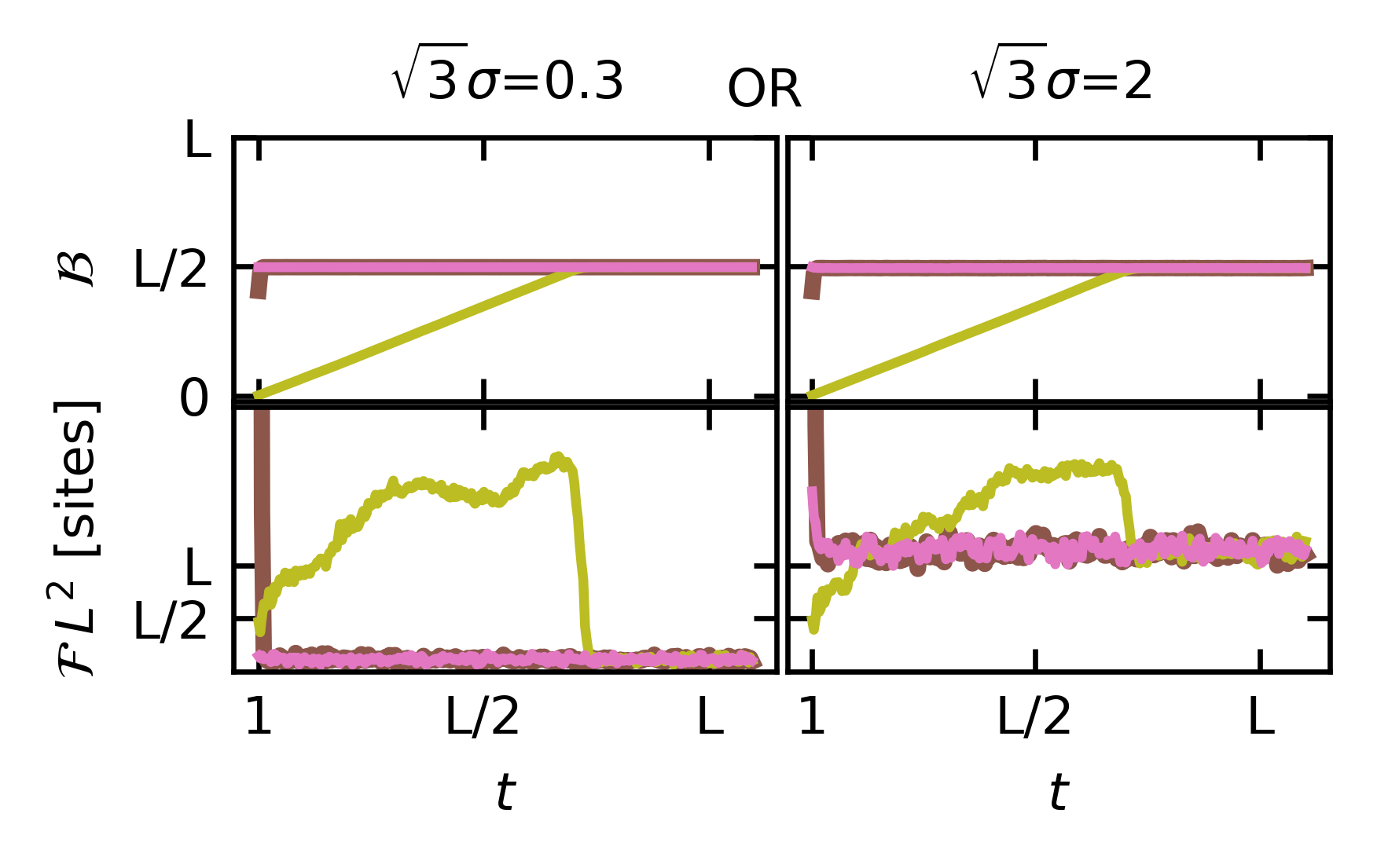}
	\caption{Exemplary time traces from simulation with same parameters as in the main text.
	For the SUM logic, we observe that the intermediate meta stable state of a half-filled boundary is realized for the boundary transition from $L/2-1$ to $L/2$ (green line).}
	\label{fig:ANDOREquivalence}
\end{figure}

The choice of $a^\text{AND}=-a^\text{OR}$ implies close correspondence of the AND and the OR rule in the stationary state.
For an infinite grid we have
\begin{equation*}
	\langle c_{ij}^\text{AND}\rangle = - \langle c_{\tilde{i}j}^\text{OR}\rangle\quad\text{with}\quad\tilde{i}= 2\frac{a}{m}-i \: ,
\end{equation*}
where $\langle \rangle$ denotes an ensemble average.
Visually speaking $\tilde{i}$ is the mirror reflection of $i$ at the zero transition of the morphogen gradient
minus its threshold at $\frac{a}{m}$.
For a finite grid, this relation still holds true for parameter combinations $a,\: m$ such that the boundary is distant from edges of the grid.
Then we can assume that cells not covered by the $\tilde{i}$ index, which are cells close to the grid boundaries, do not change their state. \\
We can derive the above relation as follows:
\begin{align*}
	c_{i,j}^\text{AND} &= \mathcal{L}^\text{AND} \left(\Lsij, \Gsij\right)\\
	&= \Theta\left(\Lsij-a^\text{AND}\right) \Theta\left(\Gsij-a\right) -\frac{1}{2} \: ,\\
\text{whereas} & \\
	-c_{\tilde{i},j}^\text{OR} &= - \mathcal{L}^\text{OR} \left(s^L_{\tilde{i}j}, s^G_{\tilde{i},j} \right)\\
	&= -\left\{1- \left[1-\Theta\left(\Lsij-a^\text{OR}\right)\right]\right.\\
	&\quad\left.\left[1-\Theta\left(s^G_{\tilde{i},j}(t)-a\right)\right]\right\}+\frac{1}{2}\\
	&= \Theta\left(-\Lsij+a^\text{OR}\right)\Theta\left(-s^G_{\tilde{i},j}(t)+a\right)-\frac{1}{2} \: ,
\end{align*}
using that $1-\Theta(x)=\Theta(-x)$.\\
Now observe that
\begin{align*}
	\Theta\left(-s^G_{\tilde{i},j}(t)+a\right) &= \Theta\left(-s^G_{\tilde{i},j}(t)+a\right)\\
	&= \Theta\left(-\tilde{i}m -\xi^G_{\tilde{i},j}+a\right)\\
	&= \Theta\left(im-\GNij-a\right) \: ,
\end{align*}
where we can neglect the sign change for $\GNij$ as it is symmetric around its zero mean.
Thus, the contribution by the global signal is by construction of $\tilde{i}$ the same as in the case of the AND rule. \\
In the stationary state, we have on ensemble average that
\begin{equation}
	\left\langle -\sum\limits_{(k,l)\in \mathcal{V}(\tilde{i},j)}c_{k,l}\right\rangle = \left\langle \sum\limits_{(k,l)\in \mathcal{V}(i,j)}c_{k,l}\right\rangle
\end{equation}
as the global signal is mirror antisymmetric with respect to the vertical $i=i_c$ line and the local signal is independent of the absolute position. \\
Thus,
\begin{align*}
	&\left\langle \Theta\left(-s^L_{\tilde{i},j}(t)+a^\text{OR}\right)\right\rangle\\
	&=\left\langle \Theta\left(-\sum\limits_{(k,l)\in \mathcal{V}(\tilde{i},j)}c_{k,l}-\xi^L_{\tilde{i},j}+a^\text{OR}\right)\right\rangle\\
	&= \left\langle \Theta\left(\sum\limits_{(k,l)\in \mathcal{V}(i,j)}c_{k,l}+\LNij-a^\text{AND}\right)\right\rangle \: .
\end{align*}
Inserting those observations yields the relation Eq.~(\ref{eq:ANDOReq}).
Exemplary time traces from simulation are shown in Fig.~\ref{fig:ANDOREquivalence}.

\FloatBarrier

\section{Time traces for even system length}\label{Supp:EvenSystemLength}
\begin{figure}
	\centering
	\includegraphics[width=1.0\linewidth]{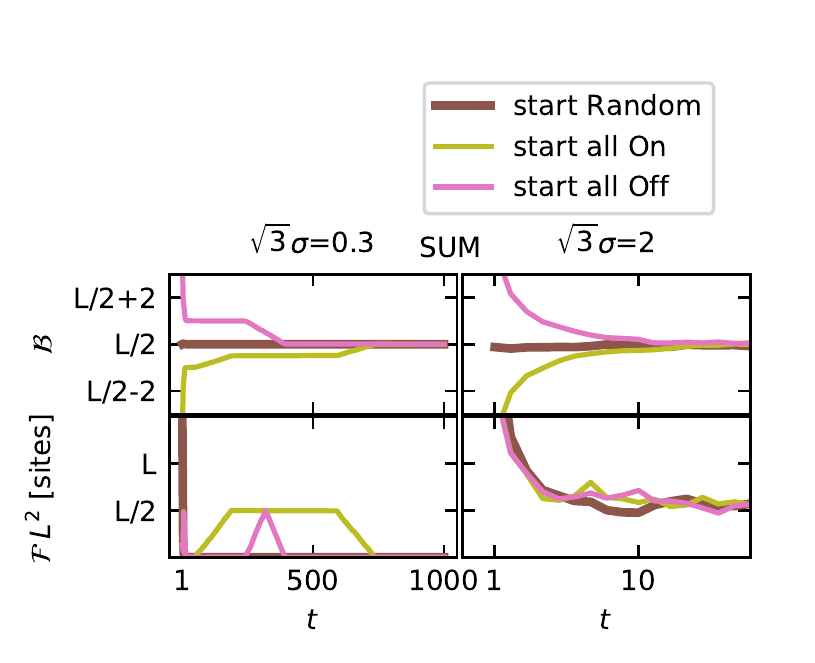}
	\includegraphics[width=0.9\linewidth]{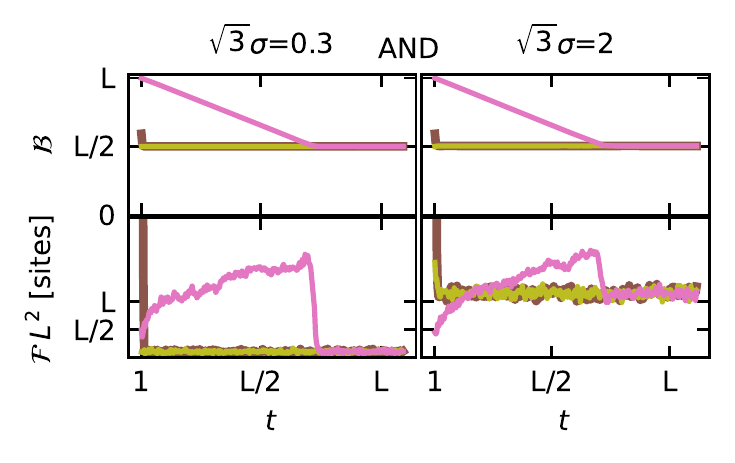}
	\caption{Exemplary time traces from simulation with same parameters as in the main text, except from the grid length, here $L=256$ sites.}
	\label{fig:timetraces_even}
\end{figure}

In Fig.~\ref{fig:timetraces_even} we show the time traces of the AND and OR logic as in Fig.~\ref{fig:timetraceplot} of the main text but for an even grid length of $L=256$ instead of the odd $L=255$.
An even grid length introduces an additional metastable state for the SUM rule between each subsequent sharp boundaries: the half-filled state, as explained in the main text. For simplicity, we thus chose to show the closest odd state grid in the main text. Here, we display the even $L=256$ version for completeness to show that except for this additional intermediate metastable state nothing changes.

\FloatBarrier

\section{Analytics for the stationary boundary position}\label{Supp:AnalyticsStationaryBP}
\paragraph{Condition for stationary boundary position}\label{par:StationaryBP}
We have two conditions to be satisfied such that the stationary boundary position is at cell index $i = i_c$:
The probability to destabilize a boundary at $i_c-1$ by an Off-in-On defect has to be smaller or equal to
the probability destabilizing a boundary at $i_c$ by an On-in-Off defect.
For a sketch, see the first grid of Fig. \ref{fig:timetraceplot}, insets (iii) and (iv), respectively.
For the right-hand side of $i_c$ we can formulate the conditions as:
\begin{align*}
	&(i) &\text{P(OffInOn $|b=i_c-1$)} &> \text{P(OnInOff $|b=i_c$)}\\
	&(ii)&\text{P(OffInOn $|b=i_c$)} &< \text{P(OnInOff $|b=i_c+1$)}
\end{align*}
with, e.g., P(OffInOn $|b=i_c-1$) denoting the probability of an Off-in-On defect, if the sharp boundary position is at $b=i_c+1$.

\paragraph{Stationary boundary position for the SUM rule}\label{par:SUMBP}
For the SUM rule the individual probabilities are given by
\begin{align*}
	\text{P(OnInOff $|b=i$)}
	&=\text{P} (\Gs(i)+\Ls(i) \geq a)\\
	&= \text{P} \left(mi+\sum\limits_{(k,l)\in \mathcal{V}(i,j)}c_{k,l} + \xi \geq a \right)\\
	&= \text{P} \left(\xi \geq a+1 -mi\right)\:,
\end{align*}
with $\mathcal{V}(i,j)$ the Von Neumann neighborhood of the cell at $(i,j)$, i.e., its upper, lower, right, and left neighbor
and $\xi = \xi^L+\xi^G$ the total noise.
We at first used that with a straight boundary at $i$ a cell at $(i,j)$ ($j$ arbitrary) has one On and three Off neighbors. \\
Further,
\begin{align*}
	\text{P(OffInOn $|b=i$)}
	&=\text{P} (\Gs(i+1)+\Ls(i+1) < a)\\
	&= \text{P} \left(m(i+1)+\sum\limits_{(k,l)\in \mathcal{V}(i+1,j)}c_{k,l} +\xi < a \right)\\
	&= \text{P} \left(\xi < a-m(i+1)-1\right)\: .
\end{align*}
Inserting both into conditions (i) and (ii) and that $\xi$ has the same distribution as $-\xi$ gives
\begin{align*}
	(i)&\quad\text{P}(\xi<a-m i_c-1) > \text{P}(\xi \leq - (a-m i_c)-1)\\
	&\Rightarrow\quad a-m i_c \geq 0 \: ,\\
	(ii)&\quad \text{P}(\xi<a-m(i_c+1)-1) < \\
	&\quad \text{P}(\xi \leq - (a-m(i_c+1))-1)\\
	&\Rightarrow\quad a-m(i_c+1) < 0 \: .\\
\end{align*}
As $i_c\in \mathbb{N}$, these two inequalities are satisfied by
\begin{equation}
	i_c^\text{SUM} = \floor{\frac{a}{m}} = i_c^\text{Grad} \: .
\end{equation}
The stationary boundary will scale with system size in the same way as for the pure boundary formation by gradient mechanism for zero noise. \\
Note that in the stationary state, implying that $i=i_c$,
the probabilities for an On-In-Off defect and an Off-In-On defect only depend on the morpophogen slope $m$
and the deviation of $\frac{a}{m}$ from its subsequent integer:
\begin{align*}
	\text{P(OffInOn $|b=i_c$)}
	&= \text{P} \left(\xi < a-m(i_c+1)-1\right)\\
	&= \text{P} \left(\xi < m\left(\frac{a}{m}-\floor{\frac{a}{m}}+1\right)-1\right)
\end{align*}
and equivalently for P(OnInOff $|b=i_c$).
\paragraph{Stationary boundary position for the AND and OR rule}\label{par:AND-OR BP}
Let us consider the AND rule.
The individual probabilities are given by
\begin{align*}
	&\text{P(OnInOff $|b=i$)} \\
	&= \text{P} \left(\Gsij+\GNij > a \right)\text{P}\left(\sum\limits_{(k,l)\in \mathcal{V}(i,j)}c_{k,l} +\LNij \geq -1 \right)\\
	&=\text{P} (\Gsij+\GNij > a)\text{P}\left(\LNij \geq 0 \right)\\
	&=\text{P} (\Gsij+\GNij > a)\frac{1}{2} \: ,
\end{align*}
where we at first used that with a straight boundary at $i$ a cell at $i$ has one On and three Off neighbors.
Then we observed that the for nonzero noise the probability of the local noise to exceed its mean 0 is 1/2,
independent of its precise distribution (as long as it is symmetric). \\
Further,
\begin{align*}
	&\text{P(OffInOn $|b=i$)} \\
	&= 1-\text{P} (\Gs_{i+1,j}+\GNij > a)\\
	&\quad\quad\cdot\text{P}\left(\sum\limits_{(k,l)\in \mathcal{V}(i+1,j)}c_{k,l} +\LNij \geq -1 \right)\\
	&=1-\text{P} (\Gs_{i+1,j}+\GNij > a)\text{P}\left(\LNij \geq -2 \right)\\
	&\approx 1-\text{P} ((\Gs_{i+1,j})+\GNij > a)
\end{align*}
Here we used that for an Off-in-On defect given the boundary is at $i$, we need to consider a cell at $i+1$,
which consequently has one Off and three On neighbors. \\
For the noise level regime, $\nl\in \left[0, 2.5\right]$ that we consider in this paper,
$\text{P}\left(\LN \geq -2 \right)\approx 1$ is a good approximation. \\
With that follows from condition (i)
\begin{equation}
	\text{P} \left(\Gs_{i_c^\text{AND},j}+\GNij > a \right) < \frac{2}{3} \: ,
\end{equation}
which is easily satisfied for $i_c^\text{Grad}=\floor{\frac{a}{m}}$ as $\Gs(i_c^\text{Grad})\approx 0$.\\
From condition (ii) follows
\begin{equation}
	\text{P} \left(\Gs_{i_c^\text{AND}+1,j}+\GNij > a \right) \geq \frac{2}{3} \: ,
\end{equation}
which consequently determines $i_c^\text{AND}$. For the Gaussian noise distribution with mean zero
and standard deviation $\Gstd = \sigma\sqrt{1-\frac{2}{2+mL}}$, it follows
\begin{align}
	i_c^\text{AND} &= \floor{\frac{a}{m}}-\frac{\nl}{m}\sqrt{1-\frac{2}{2+mL}}\left(\sqrt{\frac{2}{3}}\text{erfc}^{-1}\left(\frac{4}{3}\right) \right)\nonumber\\
	&\approx \floor{\frac{a}{m}}+0.25\frac{\nl}{m}\sqrt{1-\frac{2}{2+mL}} \: .
\end{align}
We see that $i_c^\text{AND}\approx i_c^\text{Grad} +\frac{\nl}{4m}$ for $mL\gg 2$,
which agrees nicely with simulation results shown in Fig. \ref{fig:overviewplot}. \\
For the OR rule, it follows
\begin{equation}
	i_c^\text{OR} \approx \floor{\frac{a}{m}}-0.25\frac{\nl}{m}\sqrt{1-\frac{2}{2+mL}} \: ,
\end{equation}
respectively by the AND-OR equivalence established in Eq. (\ref{eq:ANDOReq}).

\section{Convergence during Scaling} \label{Supp:ScalingConvergence}
The reason the boundary position and fuzziness for small lengths deviate from the values for large grids is of technical nature,
as our parameter choice yields a boundary at $\floor{\frac{a}{m}}=\frac{a}{m}=\frac{1}{4}L$.
For the \Grad\ rule, for instance, this choice implies
\begin{equation*}
	c_{i_c,j} = \Theta(mi_c+\GNij \geq a)-\frac{1}{2}= \Theta(\GNij\geq 0)-\frac{1}{2}\: ,
\end{equation*}
and thus on average the cell at $i_c$ is switched On every second time step due to noise.
In contrast, for $c_{(i_c+1,j)}$ these parameters more stably yield On
as
\begin{equation*}
	c_{i_c+1,j} = \Theta(m i_{c+1}+\GN_{i_c+1,j} \geq a)-\frac{1}{2}= \Theta(\GN\geq -m)-\frac{1}{2}\: .
\end{equation*}
Thus defect cells at the left of the boundary are more common than at its right.
\cleardoublepage

\end{document}


\title{Supplementary material: Robust boundary formation in a morphogen gradient via cell-cell signaling}

	\author{Mareike Bojer}
	\affiliation{Department of Physics, Technische Universit\"at M\"unchen, D-85748 Garching, Germany}

	\author{Stephan Kremser}
	\affiliation{Department of Physics, Technische Universit\"at M\"unchen, D-85748 Garching, Germany}

	\author{Ulrich Gerland}
	\affiliation{Department of Physics, Technische Universit\"at M\"unchen, D-85748 Garching, Germany}

\maketitle
\section{Boundary position and Fuzziness definition}\label{Supp:extremlyLargeNoise}
\paragraph{From exemplary grids}\label{Supp:Observables}
\FloatBarrier

\begin{figure}
	\centering
	\includegraphics[width=1\linewidth]{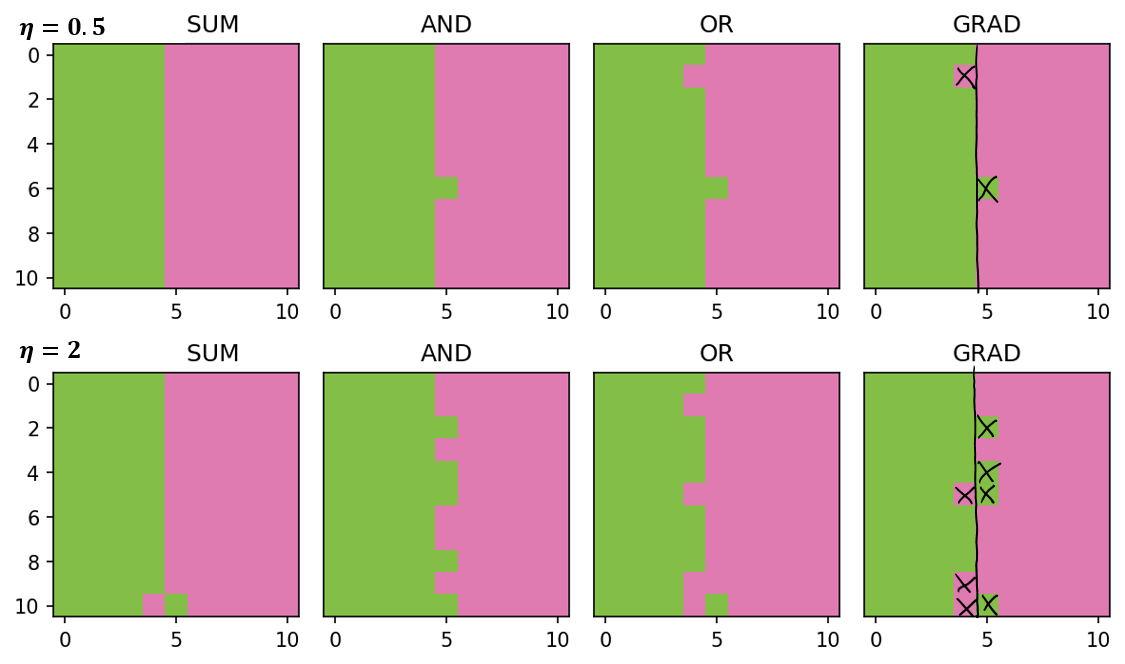}
	\caption{The first row shows an exemplary steady state grid for a small noise level of $\sqrt{2}\sigma\eta=0.5$ for each rule, the second row for a large $\sqrt{2}\sigma =\eta=2$. $L=11, \: m=1, \: a=5.5$.}
	\label{fig:gridsplot}
\end{figure}

In Fig. \ref{fig:gridsplot} we can observe in the first row for a small noise level an exemplary grid for each rule
as well as a representation for large noise in the second row.
We observe that all four rules construct the boundary at a similar position.
To make this notion quantitative,
we defined the boundary position $\mathcal{B}(t)$ of a grid at time $t$ to be the number of cells in Off state and the average over time approximating the ensemble average for sufficiently long times as formalized in the main text.
The other quality we want to quantify is the fuzziness of the boundary.
In Fig. \ref{fig:gridsplot}, for small noise levels,
the SUM rule produces a sharp boundary, the AND, OR and GRAD rule show one to two misplaced cells.
For a larger noise level, also the boundary produced by the SUM rule shows defects.
The number of defects by the AND and OR rule is similar,
although they seem to occur at different sites of the boundary.
One might even note that the number of cells in the wrong state for the AND and OR rule is smaller than for the \Grad .
A definition of boundary fuzziness is in order to discuss this heuristic hints.
The fuzziness $\mathcal{F}(t)$ of a grid at time $t$ is defined as the number of sites in the wrong state with respect to the boundary position rounded to its closest integer, also called \textit{defect cells}, in percent of the total number of cells in the grid, as stated in the main text.
In Fig. \ref{fig:gridsplot} the instantaneous fuzziness $\mathcal{F}(t)$ is visualized in terms of cells in the wrong state as crossed out cells,
for the \Grad\ rule.
For a random grid, the fuzziness reaches it's maximum value of $50\%$.\\
The ensemble average of the time it takes for a system to reach its initial condition independent state is taken to be the Transition Time $\mathcal{T}$.
Consequently the transition time is a mean first passage time.
As we are only interested in a rough estimate,
we will use the maximum of the first passage time from two runs starting at an initial grid of purely Off cells and purely On cells as an approximation.

\paragraph{Reasoning}
The definition of the boundary position definition circumvents complications due to 'holes' in the grid.
Otherwise we would have to decide to either ignore them
or introduce a left (most) and right (most) boundary and combining these in an arguable way.
Another common definition that elegantly deals with 'holes' is fitting a $\tanh$ and using the $x$-value of it's zero crossing as the boundary position. Here, we want to work with grids ranging from $2^3$ up to $2^{13}$ cells. For the small grids, fitting a $\tanh$ gives poor results.

\FloatBarrier
\section{Update scheme}\label{Supp:UpdateScheme}
\paragraph{Cellular automata simulations}\label{Supp:CASim}

Cellular automata models are models that have discretized time, state and space. 
At every time step the discrete state of each cell is updated by a function depending only on the cells and it's nearest neighbors state. 
Updates are usually synchronous. 
Here we extended the model to allow for a global signal read out at the cells position, moreover the cells future state does not directly depend on its current state.
\paragraph{Asynchronous updates yield qualitatively similar results to synchronous updates}
\label{Supp:AsyncUpdates}
\begin{figure}
	\centering
	\includegraphics[width=1\linewidth]{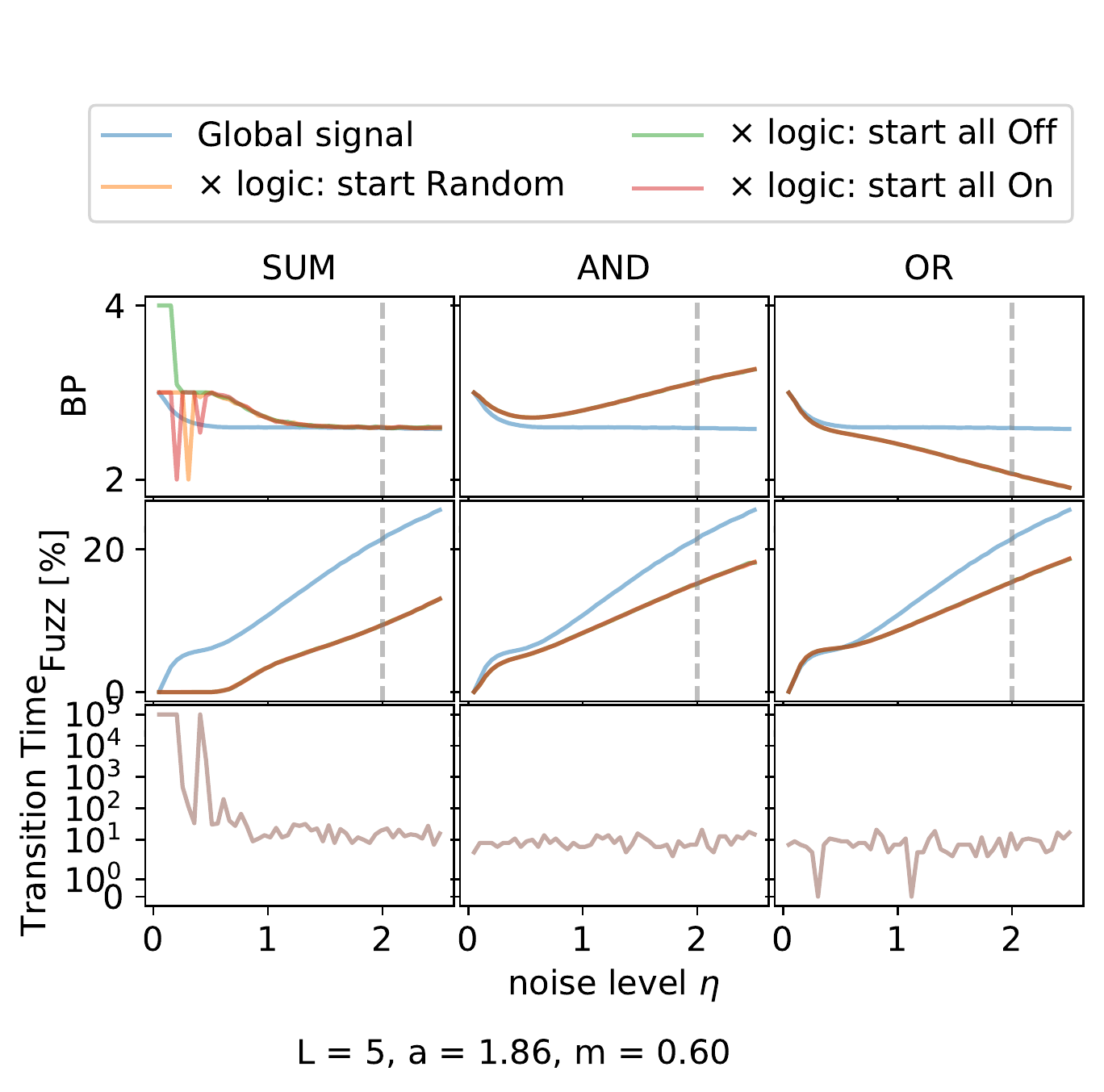}
	\includegraphics[width=1\linewidth]{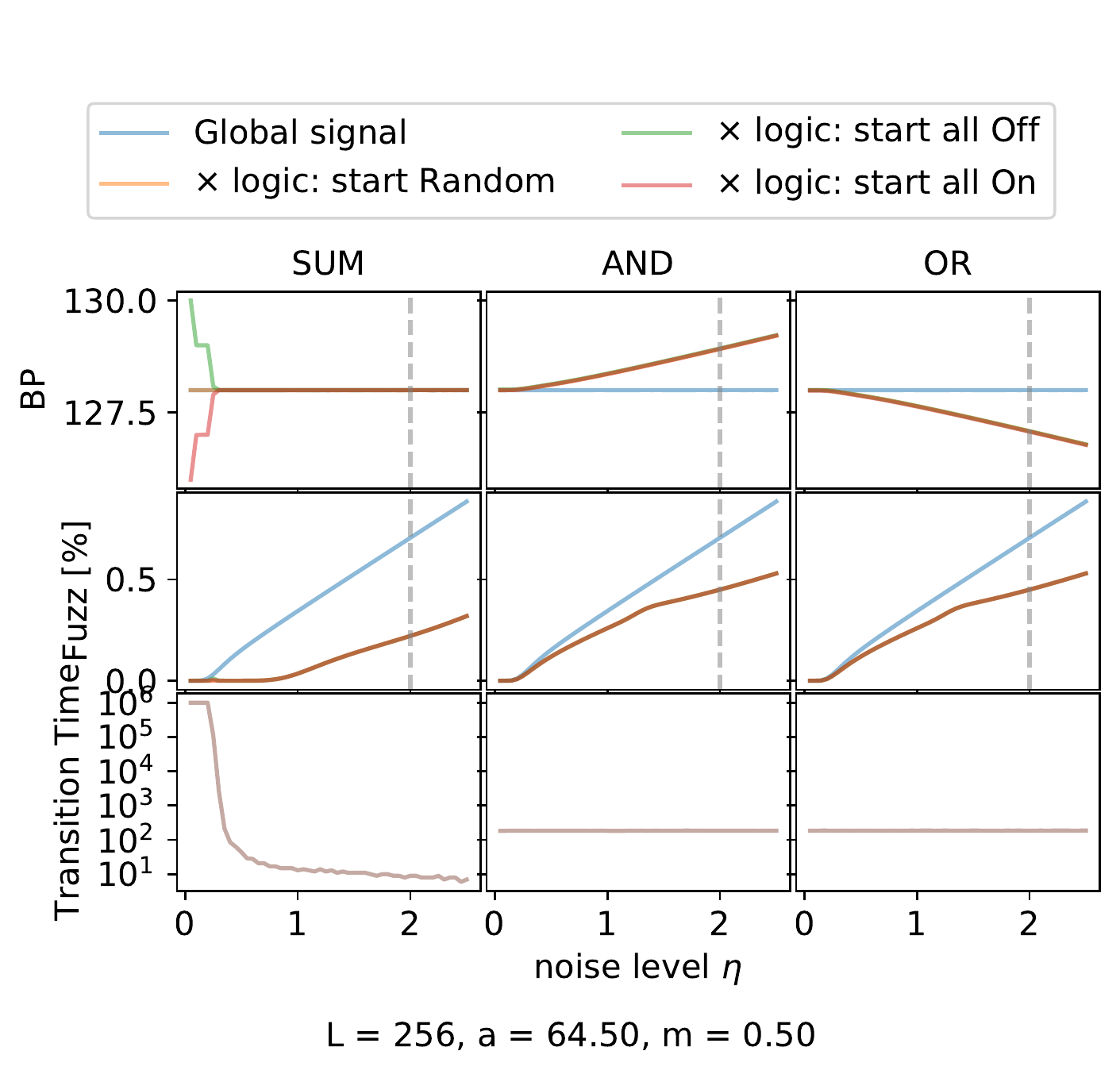}
	\caption{Upper: Overview plot for asynchronous update, Lower: Same for synchronous update}
	\label{fig:asyncoverviewplot}
\end{figure}

To make sure that our results do not depend on the exact updating procedure,
we compared the overview plot produced by the synchronous update scheme to that of a random update scheme in Fig. \ref{fig:asyncoverviewplot}.
One time step in the random update scheme corresponds to $L^2$ times drawing a cell from the grid at random and updating it according to the \Grad , SUM, AND or OR rule. \\
From Fig. \ref{fig:asyncoverviewplot}, we observe that the results qualitatively agree for all rules. Convergence for small noises is different, which makes sense as an asynchronous update could help to exit the metastable boundary position more quickly.\\
From a computational perspective the synchronous update scheme is by far more efficient than the asynchronous one.

\FloatBarrier
\section{Kinetics of the correction mechanisms in detail}\label{Supp:kineticsDetails}

\subsection*{AND rule}
When starting from an all Off initial grid,
the local signal exceeds the local threshold only at the right grid boundary
due to fixing the cells at $L+1$ to On.
The AND rule consequently conserves the Off state everywhere else.
Thus only cells at $i=L$ can switch On at the first time step.
They do so with probability one half,
as the condition of the global signal exceeding its threshold is nearly always fulfilled
and with one On neighbor the deterministic local neighbor signal equals it threshold,
thus the sign of the local noise with mean zero determines the cells state at the next time step:
\begin{align*}
	\mathcal{L}^\text{AND}\left(\Ls,\Gs\right)+\frac{1}{2}&= \Theta \left( \Lsij+1\right) \Theta\left(\Gsij -a\right) \\
	&\approx \Theta \left( \sum\limits_{j\in\mathcal{V}(i)}c_j(t)+\LNij+1\right)\cdot 1\\
	&= \Theta \left(\LNij\right)\: .
\end{align*}

If a cell has more than one On neighbor, it will probably switch on, as long as the gradient signal exceeds its threshold.
Note that both processes do not depend on the noise level, in good approximation.
As the first process repeats, noise contributions sum up,
so we observe an increase in boundary fuzziness
until the On state has propagated to its stationary boundary position.
The second process meanwhile reduces boundary fuzziness by filling wholes and is responsible for the fast convergence, once the first cell has reached to the stationary boundary position.

\subsection*{SUM rule}

\begin{figure}
	\centering
	\includegraphics[width=0.9\linewidth]{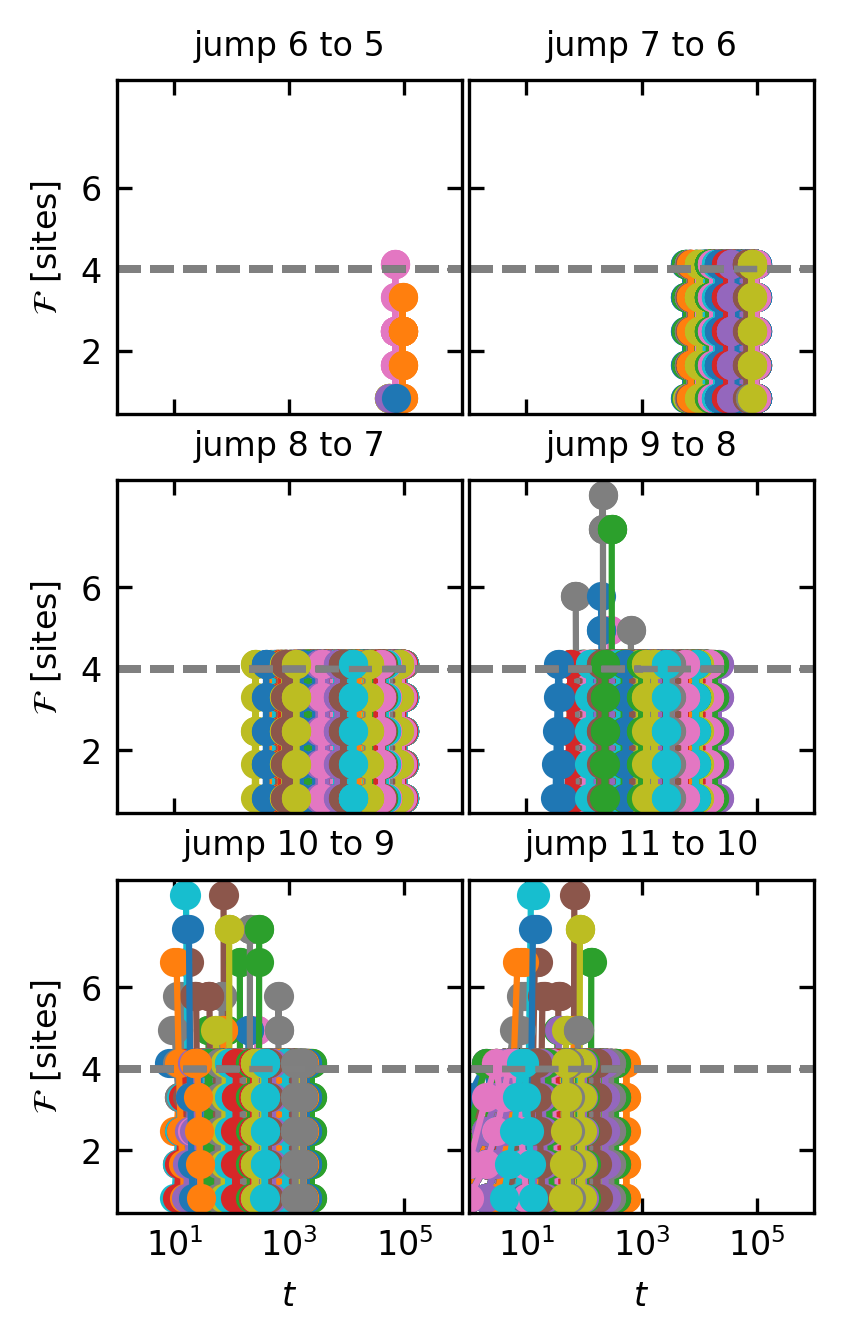}
	\caption{On an $L=11$ grid for an ensemble of 5000 runs, the fuzziness time trace in number of cells is plotted in a separate panel for each boundary jump. At $t=0$ we start from an all Off grid in the last panel and a boundary position at $L=11$ and observe that the fuzziness on average within the first 50 time steps reaches its maximum at 4, before it drops again to zero, implying that a new sharp boundary at $i=10$ has been established. Different colors indicate different runs. The full run time is limited to $10^5$ time steps, consequently in the jump 7 to 6 and 6 to 5 panel there are significantly fewer runs plotted, as for most simulations those jumps haven't happened within the time simulated. The noise level was set to $\sqrt{3}\sigma = 0.3$. }
	\label{fig:durationofboundaryhopping}
\end{figure}

At the first time step the global signal's contribution suffices to establish a boundary position close to its stationary position
and the remaining deviation is due to the different initial conditions.
Starting from a random initial grid the stationary boundary position is established so quickly that we can only observe a short peak in the fuzziness at the very beginning.
Starting from a grid of all Off (or all On cells),
we observe that the boundary position jumps very quickly from its initial value to the next straight line one cell width to its left (or right).
During the boundary jump, the fuzziness in- and decreases roughly linearly with time with a maximum value close to half of the grid's width.
This indicates that during the jump process only cells at this column change state and they do so in a consecutive, ordered manner.
From resolving this boundary jump process more accurately in time for an ensemble of runs  Fig.\ref{fig:durationofboundaryhopping},
we conclude that for sufficiently small noise
the stochastic switch of state of one cell of an otherwise straight boundary
triggers the consecutive switch of the remaining boundary cells.
This initial switch of state we refer to as a seed.
As sketched in Fig. 2 of the main text, insert (iii) and (iv), 
one seed induces a switch of both of its neighbors in the next time step and so forth until the complete column of former boundary cells has switched state, in case of an odd boundary length.

For an even grid length, half filling of the column corresponds to an intermediate meta stable state.
The next boundary jump takes significantly longer.
Boundary jumping halts when the probability for a seed destabilizing the On state is equally high to the probability of destabilizing the Off state modulo discretization.
This is the case for $i=i_c:=\floor{\frac{a}{m}}$,
which is identical to the position of the run starting from random initial conditions.

\begin{figure}
	\centering
	\includegraphics[width=1\linewidth]{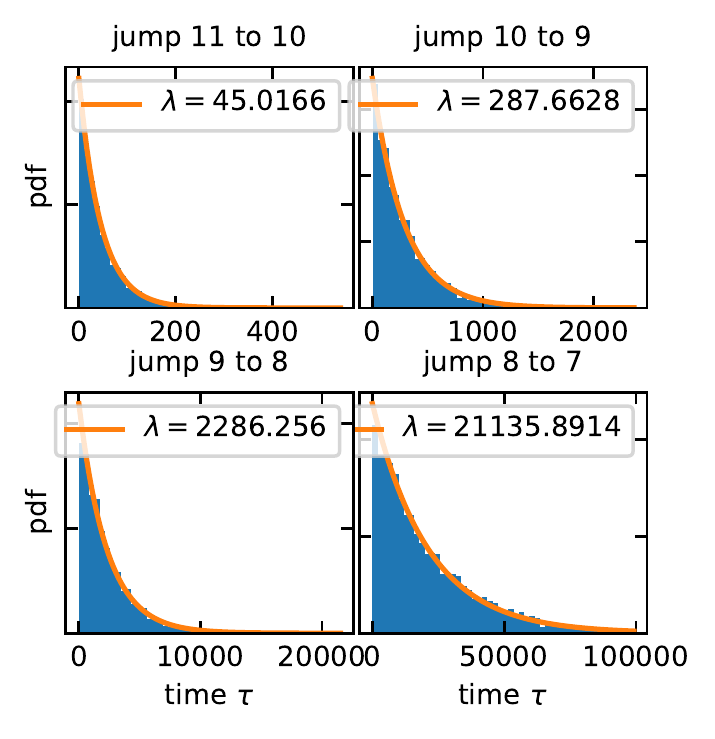}
	\caption{Shown is the waiting time distribution $\tau$ until the next boundary jump for an ensemble of 5000 runs and maximal total simulation time of $10^5$ time points. In orange the fit result for the distribution is shown, $P(\tau, i)\propto \exp{\left(-\lambda(i)\tau\right)}$, with $i$ denoting the boundary position. Results are for an $L=11$ square grid and a noise level of $\sqrt{3}\sigma=0.3$. }
	\label{fig:exponentialdecayingwaitingtimedistribution}
	\includegraphics[width=0.9\linewidth]{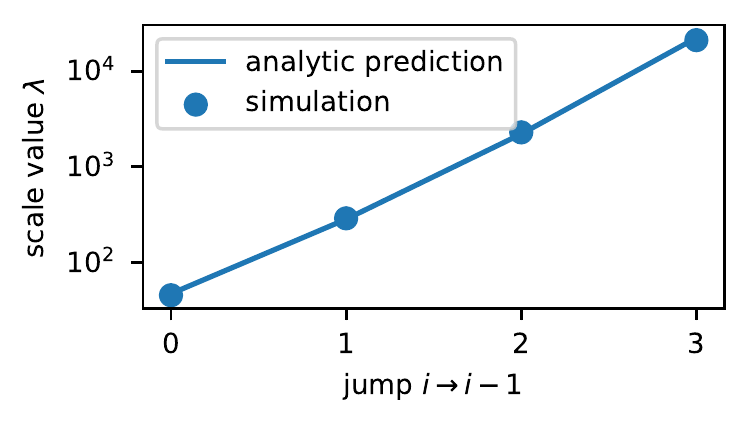}
	\caption{Comparison of the simulation results for the scale $\lambda(i)$ of each boundary jump to the analytic prediction, $\lambda^{-1}= \frac{L}{2}\text{erfc}\left(\frac{1}{\sqrt{2}}\frac{a+1-im}{\sigma}\right)$. Jumps are numbered in order of their occurrence starting from 0, i.e. jump 0 corresponds to $i=11\: \rightarrow \: i=10$, jump 1 to $i=10\: \rightarrow\: i=9$, ect. }
	\label{fig:transitiontimesmallnoise}
\end{figure}

\paragraph{Quantitative take on transition time of SUM rule}\label{par:quantiativeSUMTransitionTime}
To get a better understanding of SUM rule dynamics, we quantitatively work out the transition time for (very) small noise.
The scaling with noise level $\nl$ of the resulting transition time approximation will moreover explain,
why we cannot just simulate long enough get the SUM rules stationary state behavior in this noise regime.
\\
The probability density of the first passage time (FPT) from boundary position $i$ to $i+1$ depends on the probability of a seed for the jump, $P(\text{seed})$, as follows
\begin{align*}
	P(\text{FPT of $i\rightarrow i-1$}=t) &= (1-P(\text{seed}))^{t-1} P(\text{seed}) \\
	&= \left(\frac{1}{1-P(\text{seed})}\right)^{-t} \frac{P(\text{seed})}{1-P(\text{seed})}\\
	&\approx \lambda \exp(-\lambda t)
\end{align*}
with $\lambda = P(\text{seed})$ and using $P(\text{seed})\ll 1$.
The approximation to an exponential distribution is also confirmed numerically, see \ref{fig:exponentialdecayingwaitingtimedistribution} and Fig. \ref{fig:transitiontimesmallnoise}.

Consequently, the mean first passage time (MFPT) of the boundary jump from $i$ to $i-1$ is given by $\lambda^{-1}$.
$P(\text{seed})$ is the probability of a single cell at the boundary to switch to the wrong state,
which we call a "defect".
For an On defect at the left side of the boundary it is given by grid height $L$ times the probability that any of the boundary cells switches state.
The probability of a single cell to switch equals the cumulative Gaussian total noise distribution with standard deviation $\sigma$ for noise realizations exceeding the threshold $a$
reduced by the local signal $\Ls$ and global signal $\Gs=im$ contribution.
A cell left of the boundary has three Off neighbors and one On neighbor,
thus $\Ls=-1$.
With that
\begin{align}
	P(\text{seed}) &= L \int\limits_{a+1-im}^\infty \frac{1}{\sigma\sqrt{2\pi}}\exp\left(\frac{1}{2}\left(\frac{n}{\sigma}\right)^2\right)\d n \nonumber\\
	&=\frac{L}{2} \text{erfc}\left(\frac{1}{\sqrt{2}} \frac{a+1-im}{\sigma} \right).\label{eq:Pseed}
\end{align}
We can read off that the mean first passage time rapidly increases with decreasing $i$
thus to get an estimation of the transition time towards the stationary boundary position, we can neglect all previous boundary jumps
\begin{align*}
	\mathcal{T}(\sigma) & \approx \text{MFPT}(i_c+1 \rightarrow i_c)\\
	&\approx \frac{2}{L}\text{erfc}^{-1}\left(\frac{1}{\sqrt{2}}\frac{1}{\sigma}\right)\\
	&\approx \frac{12}{L}\exp\left(\frac{3}{2}\frac{1}{\nl^2}\right)
\end{align*}
where in the last step we have used an approximation of the complementary error function valid for small $\sigma$.
We immediately see that simulating the stationary state of very small $\sigma$ is not feasible.
\FloatBarrier

\section{Comparison to Ising}\label{Supp:Ising}
`Ising model' generally refers to an equilibrium model, whereas the CA model also describes dynamics. To nevertheless compare both models, we choose Monte Carlo updates for the Ising model and random updates (i.e. only one cell at a time) for the cellular automaton. \\
For all possible grid configurations we need to compare transition probabilities. As the interaction is local and only one site is changed at a time, it is sufficient to compare $P^\text{Ising}_{-1\rightarrow 1}$ and $P^\text{CA}_{-1\rightarrow 1}$ for a configuration $s$ such that $s_k(t)=-1$ as well as $P^\text{Ising}_{1\rightarrow -1}$ and $P^\text{CA}_{1\rightarrow -1}$ for $s_k(t)=1$. $P_{-1\rightarrow -1}$ and $P_{1\rightarrow 1}$ follow as one minus the checked ones. \\
Concretely we can pose the following question: is it possible to find a noise distribution for the cellular automaton s.t. the Ising MC-update probability $P^\text{Ising}_{-1\rightarrow 1} = P^\text{CA}_{-1\rightarrow 1}$ , can $P^\text{Ising}_{1\rightarrow -1} = P^\text{CA}_{1\rightarrow -1}$ also be fulfilled? \\
\paragraph{For the Ising model} with an MC updating scheme: Choose a random site $k$ in state $s_k$: \\
If $r\propto$Uni[0,1] $< P =$ Min$\left[1, \exp\left(- \frac{\Delta E}{T} \right) \right]:$ flip state,\\
with $k_B$ set to 1, and Uni[0,1] referring to the uniform distribution within the Interval [0, 1] and
\begin{align*}
	\Delta E &= \Delta H\left( s_k(t+1), s_k(t) \right) \\
	&= - J \sum\limits_{j\in\mathcal{N}(k)}\left(s_k(t+1)-s_k(t) \right)s_j(t)\: , \\
\end{align*}
with
\begin{align*}
	H = - J \sum\limits_{\left<i,j \right>}s_i s_j - \sum_i (gx_i + c)s_i\: .
\end{align*}
\paragraph{For the CA model} choose a random site $k$\\
If signal($k$) + noise($n$, $k$) $>a$: remain respectively switch On,\\
with signal($k$)$=\sum\limits_{j\in\mathcal{V}(k)} s_j + km$\\
$\Leftrightarrow$ if noise$(n,k)>-\sum\limits_{j\in\mathcal{V}(k)} s_j + km:\: s_k(t+1)=1$\\
A comparison to the Ising model with MC updates suggests that if there exists a probability distribution noise$(n,k)$ such that Ising corresponds to CA then it also exists for 0 external field and $J=1$ and vice versa. \\ \\
Let us start with the actual comparison. \\
Assume $s_k=-1$, then for the Ising model
\[
	P^\text{Ising}_{-1\rightarrow 1} =
	\begin{cases}\begin{aligned}
		\exp\left(-\frac{\Delta E}{T}\right) &\text{ for}&\Delta E >0\\
		= 1 &\text{ for} &\Delta E < 0
		\end{aligned}
	\end{cases}
\]
\[
 =
\begin{cases}\begin{aligned}
		\exp\left(-\frac{-2\sum_{j\in \mathcal{V}(i)}s_j}{T}\right) &\text{ for}&-\sum_{j\in \mathcal{V}(i)}s_j >0\\
		= 1 &\text{ for} &-\sum_{j\in \mathcal{V}(i)}s_j < 0
	\end{aligned}
\end{cases}
\]
In case of the CA noise($n,k$) is a random number $r,\: r\propto f(r)$ implying
\begin{align*}
	P^\text{CA}_{-1\rightarrow 1} = P\left(r >  -\sum_{j\in \mathcal{V}(i)}s_j\right) = \int\limits_{-\sum_{j\in \mathcal{V}(i)}s_j}^\infty r f(r) \d r
\end{align*}
This form suggests the choice of an exponential distribution:
\[
P^\text{CA}_{-1\rightarrow 1} =
\begin{cases}\begin{aligned}
		\exp\left(-\frac{-2\sum_{j\in \mathcal{V}(i)}s_j}{\sigma}\right) &\text{ for}&-\sum_{j\in \mathcal{V}(i)}s_j >0\\
		= 1 &\text{ for} &-\sum_{j\in \mathcal{V}(i)}s_j < 0
	\end{aligned}
\end{cases}
\]
Consequently
$P^\text{CA}_{-1\rightarrow 1}= P^\text{Ising}_{-1\rightarrow 1}$.
But as $P^\text{CA}_{-1\rightarrow 1}= P^\text{CA}_{1\rightarrow 1}$ and $P^\text{Ising}_{1\rightarrow 1} = 1 - P^\text{Ising}_{-1\rightarrow 1}$ the transition probabilities $P^\text{CA}_{1\rightarrow 1}$ and $P^\text{Ising}_{1\rightarrow 1}$ cannot coincide.
